\documentclass[structabstract]{aa}  
\usepackage{natbib}
\bibpunct{(}{)}{;}{a}{}{,}
\usepackage{graphicx}
\usepackage{aalongtable,lscape,longtable}
\usepackage{txfonts}
\begin{document}

   \title{Chromospheric activity and rotation of FGK stars in the solar vicinity\thanks{
    Based on observations made with the 2.2 m telescope at the Centro Astron\'omico Hispano Alem\'an (CAHA) at Calar Alto (Spain) 
and the Telescopio Nazionale Galileo (TNG) operated on the island of La Palma by the Istituto Nazionale de Astrofisica Italiano (INAF), 
in the Spanish Observatorio del Roque de los Muchachos. This research has been supported by the Programa de Acceso a Infraestructuras 
Cient\'{i}ficas y Tecnol\'ogicas Singulares (ICTS).} \thanks{Tables \ref{parameters} to \ref{jitter} are only available in electronic form
at the CDS via anonymous ftp to cdsarc.u-strasbg.fr (130.79.128.5) or via http://cdsweb.u-strasbg.fr/cgi-bin/qcat?J/A+A/}}

   \subtitle{An estimation of the radial velocity jitter}

   \author{R. Mart\'{i}nez-Arn\'aiz \inst{1},  
          J. Maldonado \inst{2},
          D. Montes \inst{1},
          C. Eiroa \inst{2},
          B. Montesinos \inst{3}
%    \fnmsep\thanks{Just to show the usage
%          of the elements in the author field}
          }
   \institute{Universidad Complutense de Madrid,
              Facultad de Ciencias F\'{i}sicas, Dpt. Astrof\'{i}sica, av. 
              Complutense s/n. 28040 Madrid, Spain \\
              \email{rma@astrax.fis.ucm.es}
         \and
             Universidad Aut\'onoma de Madrid,
             Departamento de F\'isica Te\'orica, M\'odulo 15, 28049 Cantoblanco, Madrid, Spain
%             \email{}
          \and 
             LAEX, CAB (CSIC-INTA), ESAC Campus, P.O. BOX 78, 28691 Villanueva de la Ca\~nada, Madrid, Spain
%             \thanks{The university of heaven temporarily does not
%                     accept e-mails}
             }            
\offprints{R. Mart\'inez-Arn\'aiz (rma@astrax.fis.ucm.es)}
%   \date{Received September 15, 1996; accepted March 16, 1997}

% \abstract{}{}{}{}{} 
% 5 {} token are mandatory
 
  \abstract
   {Chromospheric activity produces both photometric and spectroscopic variations that can be mistaken as planets. Large spots crossing the 
stellar disc can produce planet-like periodic variations in the light curve of a star. These spots clearly affect the spectral line 
profiles and their perturbations alter the line centroids creating a radial velocity jitter that might ``contaminate'' the variations 
induced by a planet. Precise chromospheric activity measurements are needed to estimate the activity-induced noise that should be expected 
for a given star.}
   {We obtain precise chromospheric activity measurements and projected rotational velocities for nearby 
(d $\leq$ 25 pc) cool (spectral types F to K) stars, to estimate their expected activity-related jitter. 
%These stars are the ``natural'' targets of exoplanet searches and thus they must be well characterized in terms of intrinsic $RV$ variations.\\
As a complementary objective, we attempt to obtain relationships between fluxes in different activity indicator lines, that permit 
a transformation of traditional activity indicators,  $i$.$e$, Ca \textsc{ii} H \& K lines, to others that hold noteworthy advantages.}
%Although the red region of cool stars spectrum present several advantages when searching for exoplanets (higher S/N ratio, lower effect of spots 
%in photometric and spectroscopic variations), chromospheric activity has been traditionally studied using the blue region, $i$.$e$, 
%the Ca \textsc{ii} H \& K lines. By building empirical relationships between them and other chromospheric activity indicator lines that appear 
%at larger wavelengths, these advantages could be exploited.}
   {We used high resolution ($\sim$ 50000) $echelle$ optical spectra. Standard data reduction was performed using the IRAF \textsc{echelle} package. 
To determine the chromospheric emission of the stars in the sample, we used the spectral subtraction technique. We measured the equivalent widths 
of the chromospheric emission lines in the subtracted spectrum and transformed them into fluxes by applying empirical equivalent width and flux 
relationships. 
Rotational velocities were determined using the cross-correlation technique. To infer activity-related radial velocity ($RV$) jitter, 
we used empirical relationships between this jitter and the $R'_{\rm HK}$ index.}
   {We measured chromospheric activity, as given by different indicators throughout the optical spectra, and projected rotational 
velocities for 371 nearby cool stars. We have built empirical relationships among the most important chromospheric emission lines. Finally, 
we used the measured chromospheric activity to estimate the expected $RV$ jitter for the active stars in the sample.}
   {}

   \keywords{Galaxy: solar neighbourhood -- Stars: late-type -- Stars: activity -- Stars: chromospheres -- Stars: rotation 
                              -- Stars: planetary systems}
   \authorrunning{Mart\'{i}nez-Arn\'aiz et al.}
   \titlerunning{Chromospheric activity and rotation of FGK stars in the solar vicinity.}
   \maketitle
%
%________________________________________________________________

\section{Introduction}
Exoplanetary science is living a golden era characterized by the enormous rate at which new exoplanets are being discovered. 
This impressive advancement would not have been possible without parallel technological developments. 
Improved precision in both radial velocity and photometric measurements has extended the area around the star in which a 
planet can be found and has increased the probability of detecting low mass exoplanets. These improvements have created, 
however, a new and noteworthy problem, $i$.$e$., the possibility of misidentifying planets. Chromospheric activity, 
in particular, the presence of spots and/or alteration of the granulation pattern in active regions, create a time-variable 
photometric and spectroscopic signature \citep{1997ApJ...485..319S,2000A&A...361..265S,2003ASPC..294...65S} that might be misinterpreted 
as an exoplanet. Moreover, the minimum detectable mass of a planet orbiting a star is limited by the $rms$ velocity jitter caused by 
stellar sources \citep{2005ApJ...620.1002N}. Therefore, a thorough analysis of activity levels, as only different activity indicators 
can provide, must be performed for those stars that constitute the natural targets of planet-search surveys.
 
When searching for exoplanets using the radial velocity ($RV$) technique, the possibility of 
jitter caused by chromospheric activity must be considered. When modeling stellar activity, \citet{1997ApJ...485..319S} 
were the first to quantitatively estimate the impact of stellar spots on the $RV$ curve and showed that they could 
produce peak-to-peak $RV$ amplitudes of up to a few hundred \,m\,s$^{-1}$, depending on spot size and the rotational velocity of 
the star. Subsequent studies \citep{2000A&A...361..265S,2003csss...12..694S, 2004AJ....127.3579P, 2007A&A...473..983D} 
obtained similar results. Several attempts to reduce $RV$ noise levels by using an activity-based correction have been made 
\citep{1997ApJ...485..319S,1998ApJ...498L.153S,2000A&A...361..265S,2000ApJ...534L.105S,2003csss...12..694S,2005PASP..117..657W}. 
To use these corrections and to test and calibrate the relations, high precision, homogeneous chromospheric activity measurements are needed. 

Transit searches for exoplanets are also affected by the temporal evolution and modulation of active regions across the stellar disc. 
The amplitude of variations can reach more than 1$\%$ when a large spot crosses the solar disc at activity maximum. This decrease in signal, 
combined with the modulation caused by stellar rotation can mimic the signal of a planet orbiting the star. Therefore Sun-like 
variability, not to mention that of stars more active than the Sun, can significantly affect the detection performance of photometric 
planet searches \citep{1997ApJ...474..503H,1997ApJ...474L.119B,2000ApJ...531..415H,2004A&A...414.1139A}. 

Hence, chromospheric activity is a {\itshape proxy} for predicting variability levels expected for a star, therefore 
allowing the estimation of a lower limit for planet detections in its vicinity. In this paper, 
we present spectroscopic-based chromospheric activity measurements for 371 nearby (d $\leq$ 25 pc), cool (spectral types F to K) stars. 
These stars are the natural targets for exoplanet searches: their proximity ensures the ability to get an adequate signal-to-noise 
ratio (hereafter S/N), and solar-like stars are more likely to host so-called habitable planets \citep{1993Icar..101..108K,1998Acas,
2003ApJS..145..181T}. In this study, we considered not only the traditional chromospheric activity indicators, $i$.$e$.,  
$R'_{\rm HK}$ or H$\alpha$, but also other less common indicators, such as the Ca \textsc{ii} IRT lines, which allow us to exploit 
peak-to-peak $RV$ amplitude variations caused by spots being less significant at longer wavelengths \citep{2007A&A...473..983D,arXiv:0909.0002v1}. 
\section{The stellar sample}
\label{s_sample}
The stellar sample comprises 371 cool stars in the solar vicinity, constrained to be at distances closer than 25 pc (see Table 
\ref{parameters}). Distances were obtained from the Hipparcos Catalogue \citep{HIPCatalogue} and the {\itshape New Reduction of the Raw Data} 
\citep{0067-0057}. The spectral type distribution of the sample is 56 F type stars, 126 G type stars, 186 K stars, and 3 M type stars. 
Since our study is based on solar-like stars, only F, G and K spectral types 
stars with trigonometric parallax $\pi$ $\geq$ 40 mas were retrieved. To avoid misclassified stars, we used colour index 
as a complementary criteria, $i$.$e$., $B$--$V$ colour index in the range 0.25-0.58 mag (F stars), 0.52-0.81 mag 
(G stars), and 0.74-1.40 mag (K stars). To establish the main-sequence character of the stars, we used a cutoff of $\pm$
1 mag from the main sequence \citep{2005PASP..117..657W}. Finally, stars in multiple systems were removed from the sample. 
We used the CCDM \citep{1994CoORB.115.....D,2002yCat.1274....0D} and SB9 \citep{2004A&A...424..727P} catalogues to identify astrometric and 
spectroscopic binaries, respectively. 

These stars are potential targets for present and future projects that aim to detect Earth-like planets or exo-solar analogues to the 
Edgeworth-Kuiper Belt. In this context, most of our stars will be observed in the framework of DUNES (DUst around NEarby Stars), an approved 
Herschel Open Time Key Project with the aim of detecting cool faint dusty disks, at flux levels as low as the Solar EKB. Some preliminary 
results can be found in \citet{2009iau258_rma}, \citet{2010pathways_jmp}, and \citet{2010pathways_dmg}.
\section{Observations and data reduction}
The present work is based on data extracted from high resolution {\itshape echelle} spectra. Most of the spectra 
were obtained in several observing runs but we also used $S^{4}N$ spectra \citep{2004A&A...420..183A}. The latter 
have similar spectral resolution ($\sim$ 45000) to those obtained by us and were consequently used in an analogous manner 
to measure and analyse the stellar parameters and properties of the stars.

Observations were obtained at two different observatories: the German-Spanish Astronomical Observatory, 
CAHA, (Almer\'{i}a, Spain) and La Palma Observatory (La Palma, Spain). Observations were taken at the former.2 m 
telescope using the {\itshape Fibre optics Cassegrain Echelle Spectrograph} (FOCES) \citep{1998A&AS..130..381P} with a 
2048$\times$2048 24 $\mu$m SITE$\sharp$1d-15, and at the latter with the 3.5 m Telescopio Nazionale Galileo (TNG) 
using the {\itshape Spectrografo di Alta Resoluzione Galileo} (SARG), the grid R4 (31.6 lines/mm), the red cross-dispersor 
(200 lines/mm), and a CCD detector mosaic of total surface 2048$\times$4096 and 13.5 $\mu$m pixels.
 
We carried out four observing runs at CAHA (July 2005, January 2006, December 2006 and February-May 2007). FOCES spectra have 
a wavelength range from 3600 to 10700 \AA\ in 106 orders with a typical resolution of 40000 (reciprocal dispersion from 0.08-0.13 \AA/pixel 
in the red and blue region of the spectrum, respectively). The total number of stars observed using this spectrograph is 198.
%and the spectral resolution determined as the full width at half maximum (FWHM) at the arc comparison lines, ranges from 0.08-0.40 \AA\.
At La Palma Observatory, we performed three observing runs (February 2006, April 2007 and, November 2008). SARG spectra have a wavelength range 
from 5540 to 7340 \AA\ in 50 orders with a resolution of 57000 (reciprocal dispersion from 0.01 to 0.04 \AA/pixel in the red 
and blue region of the spectrum, respectively). We observed 129 stars using the SARG spectrograph.
%and the spectral resolution (FWHM) from 0.07 to 0.16 \AA.

As mentioned before, we used $S^{4}N$ \citep{2004A&A...420..183A} spectra. These spectra were taken 
between October 2000 and November 2001 in six observing runs at the Harlan J. Smith 2.7 m telescope (McDonald Observatory) and two at 
the 1.52 m telescope at La Silla (Chile). At McDonald Observatory, the 2dcoud\'e  \citep{1995PASP..107..251T} spectrograph with 
the Tektronix 2048$\times$2048 24 $\mu$m CCD detector was used. The spectral coverage was 3600-5100 \AA\, with a typical 
resolution of 50000. At La Silla Observatory, the {\itshape Fiber-fed Extended Range Optical Spectrograph} (FEROS) 
\citep{2000SPIE.4008..459K} with the CCD detector EEV 2048$\times$2048 15 $\mu$m was used. The wavelength coverage 
in this case is 3500-9200 \AA\, and the resolution $\sim$ 45000. The total number of studied $S^{4}N$ stars is 106. Of them, 79 
were observed at McDonald Observatory, while the rest were observed with FEROS at La Silla Observatory.

All the observed stars and the corresponding spectrograph used are listed in Table \ref{parameters}. We note that some 
stars were observed more than once and using different spectrographs.

%We also used data obtained by \citep[][in prep.]{2009Lopez-Santiago}.  
For the reduction, we used the standard procedures in the IRAF ({\itshape Image Reduction and Analysis Facility}) package 
(bias subtraction, extraction of the scattered light produced 
in the optical system, division by the normalized flat-field, and wavelength calibration). After the reduction process, the spectrum 
was normalized to the continuum order by order by fitting a polynomial function to remove the general shape of the aperture spectra. 
\section{Analysis and results}
%
%\subsection{Color index}
%
%Both, the method used to compute rotational velocities (see \S \ref{vsini}) and effective temperatures (see \S \ref{s_spectral_types}) 
%required the use of colour index $B$--$V$. This index (listed in Table \ref{parameters}) was obtained from Tycho-2 Catalogue 
%\citep{2000A&A...355L..27H} using the transformation from $B_{\rm T}$ and $V_{\rm T}$ to Johnson indices (see section 1.3 from 
%Hipparcos Catalogue, ESA 1997).
% 
\subsection{Rotational velocities}
\label{s_vsini}
The determination of rotational velocities for stars in exoplanet surveys is crucial. Radial velocity variations induced 
by chromospherically active regions on the stellar surface are modulated by the rotation period of the star \citep{1997ApJ...474L.119B,
1997ApJ...474..503H,2000ApJ...531..415H}. Obtaining the rotational velocity of the star is thus essential to test whether
detected $RV$ variations have a stellar or a planetary origin.

%The determination of rotational velocities when handling precise radial velocity measurements is crucial. 
%Given that the 
%radial velocity is determined by measuring the position of the centre of the spectral lines, any process producing a modification of their shape
% will have an effect on the measured values. Stars with high rotational velocities present a non-gaussian profile and the precise determination
% of the radial velocity becomes difficult. Moreover, radial velocity variations induced by chromospherically active regions on the stellar surface 
%are modulated with the rotation period of the star. Obtaining the rotational velocity of the star is thus essential to test whether 
%detected $RV$ variations have a stellar or a planetary origin.
%
%
\begin{table*}
\caption[A constant]{A constant for FOCES, 2dcoud\'e (McDonald), FEROS and SARG spectra.}
\label{tab:A_const}
\centering
\begin{scriptsize}
\begin{tabular}{l c c l c c l c c l c c}
\hline
\hline
\multicolumn{3}{c}{FOCES} & \multicolumn{3}{c}{2dcoud\'e} & \multicolumn{3}{c}{FEROS} & \multicolumn{3}{c}{SARG}\\
\noalign{\smallskip}
%\cline{2-3} \cline{6-7} \cline{10-11} \cline{12-13}
{Name} & {SpT} & {A} &
{Name} & {SpT} & {A} &
{Name} & {SpT} & {A} &
{Name} & {SpT} & {A} \\
\hline
\noalign{\smallskip}
HIP 104217 & K7V  & 0.500 $\pm$ 0.130 & HIP 67422 & K2V & 0.550 $\pm$ 0.009 & HIP 99461 & K3V & 0.640 $\pm$ 0.015 & HIP 98698 & K4V & 0.523 $\pm$ 0.007\\
HIP 117779 & K5V  & 0.436 $\pm$ 0.095 & HIP 3765  & K2V & 0.610 $\pm$ 0.008 & HIP 10138 & K1V & 0.620 $\pm$ 0.008 & HIP 3765  & K2V & 0.605 $\pm$ 0.005\\
HIP 3765   & K2V  & 0.616 $\pm$ 0.100 & HIP 7981  & K1V & 0.580 $\pm$ 0.008 & HIP 99825 & K0V & 0.600 $\pm$ 0.008 & HIP 88972 & K2V & 0.501 $\pm$ 0.003\\
HIP 7981   & K1V  & 0.574 $\pm$ 0.100 & HIP 3093  & K0V & 0.570 $\pm$ 0.008 & HIP 84720 & G8V & 0.640 $\pm$ 0.008 & HIP 3093  & K0V & 0.607 $\pm$ 0.006\\
HIP 3093   & K0V  & 0.557 $\pm$ 0.100 & HIP 56997 & G8V & 0.630 $\pm$ 0.016 & HIP 57443 & G5V & 0.660 $\pm$ 0.008 & HIP 64924 & G5V & 0.606 $\pm$ 0.010\\
HIP 95319  & G8V  & 0.648 $\pm$ 0.100 & HIP 10798 & G5V & 0.620 $\pm$ 0.008 & HIP 91438 & G5V & 0.650 $\pm$ 0.008 & HIP 23835 & G4V & 0.646 $\pm$ 0.008\\
HIP 9269   & G5V  & 0.581 $\pm$ 0.224 & HIP 64924 & G5V & 0.600 $\pm$ 0.008 & HIP 71683 & G2V & 0.640 $\pm$ 0.008 & & & \\
HIP 48113  & G0.5V& 0.680 $\pm$ 0.280 & HIP 7918  & G1V & 0.640 $\pm$ 0.008 & HIP 1599  & F9V & 0.680 $\pm$ 0.007 & & & \\ \hline
\noalign{\smallskip}
\multicolumn{3}{c}{$\langle A \rangle$ = 0.574 $\pm$ 0.004} &
\multicolumn{3}{c}{$\langle A \rangle$ = 0.598 $\pm$ 0.003} &
\multicolumn{3}{c}{$\langle A \rangle$ = 0.643 $\pm$ 0.003} &
\multicolumn{3}{c}{$\langle A \rangle$ = 0.581 $\pm$ 0.082}\\
\noalign{\smallskip}
\hline
\end{tabular}
\end{scriptsize}
\end{table*}
The widths and shapes of spectral lines contain information that allow us to deduce physical information about the star,
including its rotation rate. To measure this width, the most commonly used parameter is the the line FWHM (Full Width Half Maximum) because 
it can be easily measured. The Fourier domain offers, however, some advantages over the wavelength one. Signatures of certain physical 
processes, such as the Doppler-shift distribution produced by rotation or macroturbulence, are more readily detected in that domain. 
The cross-correlation function (CCF) was therefore used to measure rotational velocities. 
The width ($\sigma$) of the CCF peak for a star when correlated with itself depends on the instrumental 
profile and several broadening mechanisms such as gravity, effective temperature, or rotation. To measure the rotational contribution, 
and hence determine the star's projected rotational velocity, the contribution of other broadening mechanisms has to be modelled. For 
$v$~$\sin{i}$ $\leq$ 50 \,km\,s$^{-1}$, the CCF is well approximated by a Gaussian \citep{1989AJ.....97..539S} and consequently the 
rotational broadening corresponds to a quadratic broadening of the CCF. In that case, the observed width of the CCF can be written as 
\citep[see][and references therein]{1998A&A...335..183Q}
\begin{equation}
 \sigma_{\rm obs}^{\rm 2}=\sigma_{\rm rot}^{\rm 2}+\sigma_{\rm 0}^{\rm 2},
\end{equation}
where $\sigma_{\rm rot}$ is the rotational broadening and $\sigma_{\rm 0}$ is the width of the CCF of a similar non-rotating star.
Projected rotational velocities, v~$\sin{i}$, can be derived from the above expression as
\begin{equation}
 v\,\sin\,i=A(\sigma_{\rm obs}^{\rm 2}-\sigma_{\rm 0}^{\rm 2})^{\rm 1/2},
\end{equation}
where $A$ is a coupling constant that depends on the spectrograph and its configuration. To determine $A$ for each spectrograph,
non-rotating stars were used\footnote{Slow rotating stars with well known $v$~$\sin{i}$.}. Their spectra were broadened from 
$v$~$\sin{i}$=1 \,km\,s$^{-1}$ to 50 \,km\,s$^{-1}$ using the program \textsc{jstarmod}\footnote{\textsc{jstarmod} is a modified 
version of the Fortran code \textsc{starmod} developed at the Penn State University \citep{1984BAAS...16..510H,1985ApJ...295..162B}. 
The modified code, implemented by J. L\'opez-Santiago, admits as input $echelle$ spectra obtained with a CCD with more than 2048 pixels.}. 
The value of $\sigma_{\rm obs}$ was then 
determined as the equivalent width of the first peak in the CCF. 
The constant $A$ was found by fitting the relation  $v$~$\sin{i}^{\rm 2}$ {\itshape vs}
$\sigma_{\rm obs}^{\rm 2}$. The stars used to compute $A$, the measurements, and the mean values for each spectrograph are shown
in Table \ref{tab:A_const}.

\begin{figure}[h!]
\begin{center}
\includegraphics[width=9.cm, keepaspectratio]{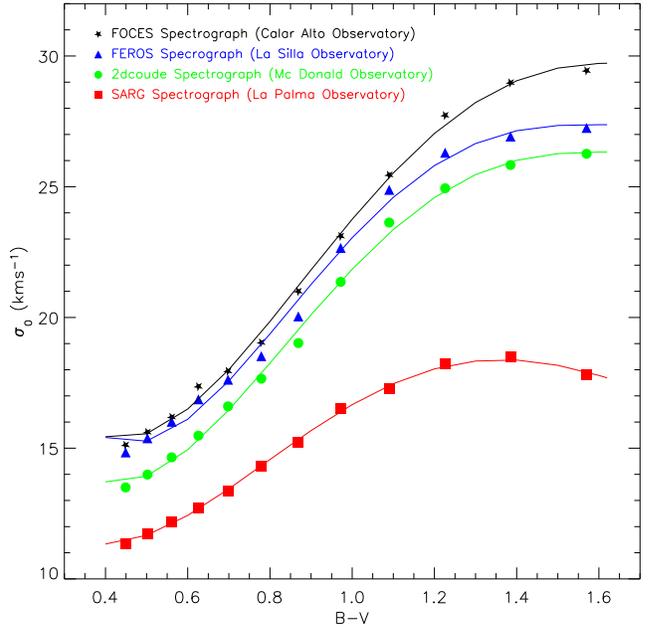}
\caption{Calibration between $\sigma_{\rm 0}$ and colour index $B$--$V$, where $\sigma_{\rm 0}$ represents the ``natural'' broadening of the spectrum lines 
and was obtained from synthetic spectra covering the spectral range of the observed stars. Given that the instrumental profile 
also contributes as a broadening mechanism and considering that each spectrograph has a different one, we had to derive a calibration 
relationship for each of them.}
\label{sigma_0}
\end{center}
\end{figure}

It is well known that $\sigma_{\rm 0}$ is a function
of the broadening mechanisms present in the atmosphere of the star, except rotation
\citep{2004IAUS..215..455M}. Since the broadening mechanisms are a function
of the temperature and gravity, we may expect  the colour index, $B$--$V$ to depend on $\sigma_{\rm 0}$, for stars
in the main sequence. To determine this dependence, we created morphed spectra with no rotational velocity using the 
ATLAS9 code by \citet{1993ASPC...44...87K} adapted to operate on a linux platform by \citet{2004MSAIS...5...93S} and 
\citet{2005MSAIS...8...61S}. Temperatures vary from 3500 K to 6500 K (in intervals
of 250 K). Since the stars were selected to be on the main sequence, $\log g$ was fixed to be 4.5. Solar metallicity
was assumed. The morphed spectra were broadened to match the instrumental
profile of the real spectra using the FWHM of the calibration arc lines, which is a good approximation of 
the broadening for the instruments used. 
Given that each spectrograph has a different resolving power, different instrumental broadenings were applied to obtain a calibration curve for each instrument. 
These curves are shown in Fig. \ref{sigma_0}.

Color indices, $B$--$V$, were obtained from the Tycho-2 Catalogue \citep{2000A&A...355L..27H} using the transformation from 
$B_{\rm T}$ and $V_{\rm T}$ to Johnson indices (see Sect. 1.3 from Hipparcos Catalogue, ESA 1997) and are listed in 
Table \ref{parameters}.

Once $A$ and $\sigma_{\rm 0}$ were known for each star, $v$~$\sin{i}$ could be directly derived by measuring $\sigma_{\rm obs}$, $i$.$e$., 
width of the CCF of the star when correlated with itself. Values for stars in the sample can be found in Table \ref{parameters}. For slow rotating 
stars, it is important to mention that sometimes, the value of $\sigma_{\rm 0}$ is larger than that of $\sigma_{\rm obs}$. 
In that case, $v$~$\sin{i}$ cannot be measured using this method and we only provide an upper limit. This value was chosen by 
considering the minimum $v$~$\sin{i}$ that could be measured with the same spectrograph for a star of the same spectral type, $i.e.$, 
the same $\sigma_{\rm 0}$.

\subsection{Chromospheric activity}
\label{s_activity}

We analyse different activity indicators throughout the optical spectra (from \ion{Ca}{ii} H\&K 
to \ion{Ca}{ii} IRT). These lines form at different heights in the chromosphere hence provide information about different 
stellar properties. Some lines are present only during high energy processes such as flares. As shown in previous 
work \citep[see][and references therein]{2000A&AS..146..103M,2001A&A...379..976M}, by performing a simultaneous analysis of 
different optical chromospheric activity indicators, a detailed study of the chromosphere's structure can be achieved and 
it becomes possible to discriminate between structures such as plages, prominences, flares, and microflares. 
The spectra used in this work have a spectral range that covers plages from \ion{Ca}{ii} H \& K to \ion{Ca}{ii} IRT lines, 
including the Balmer lines.
\subsubsection{Equivalent widths and fluxes}
\label{s_ew_flux}
To determine the chromospheric contribution to the spectrum, and thus the chromospheric activity, 
the contribution of the photosphere must be removed. In order to achieve this, we used the spectral subtraction 
technique, described in detail by \citet{1995A&AS..114..287M,2000A&AS..146..103M}. 
This technique has been extensively used before because it permits the detecion of weak emission features in the cores
of chromospheric lines. In addition, it is the most effective means of identifying other 
chromospheric activity indicators such as the Balmer lines or the \ion{Ca}{ii} IRT, where no calibrations of the photospheric 
minimum flux exists \citep{1985ApJ...295..162B,1989AJ.....98.1398H,1992AJ....104.1942H,1994A&A...284..883F,1997A&A...318...60G,
1997AJ....113.2283L,1995A&AS..114..287M,1996A&A...312..221M,1997A&AS..123..473M,2000A&AS..146..103M,2001A&A...379..976M,
2002A&A...389..524G,2007A&A...472..587G,2009AJ....137.3965G,2003A&A...411..489L,2009A&A...survey}.
%{\bfseries This technique has been proved to be equivalent to the tradicional method to determine R'$_{\rm HK}$ by many authors. 
%In addition, it is the best method to obtain other chromospheric activity indicators like the Balmer lines or the \ion{Ca}{ii} IRT 
%where no calibrations of the photospheric minimum flux exists .}
Inactive, slowly rotating stars, observed in the same observing run as the active stars, were used as reference to construct a 
morphed spectrum for each active star, using the program \textsc{jstarmod}. The program builds the morphed spectrum by 
shifting and broadening the reference spectrum to match that of the  target star. This implies that the reference star must have a 
lower rotation rate than but a similar spectral type to the target star. A compilation of the inactive, slowly rotating stars 
used as references can be found in Table \ref{ref_activity}. Reference stars were initially chosen from the literature but some
inactive, slowly rotating stars of the sample were also used as reference after confirming that they did not show chromospheric activity.
To ensure that those stars were inactive, their values of total flux in \ion{Ca}{ii} H \& K were compared to the lower boundary 
defined by \citet{1984A&A...130..353R}, which is traditionally used to correct flux measurements from the basal chromospheric flux (see Fig. 
\ref{minimun_surf_flux}). Maximum differences of 0.25 (F type stars), 0.3 (G type stars) and 0.35 (K type stars) in $\log (F_{\rm H} + F_{\rm K})$ 
values between the stars used as references and the lower boundary of \citet{1984A&A...130..353R} were allowed. 
The  morphed spectrum was subtracted from that of the active star, obtaining a spectrum in which only the chromospheric 
contribution is present, $i$.$e$., the subtracted spectrum. The excess emission $EW$ of the activity indicator lines were obtained 
from that spectrum. To estimate the errors in the measured $EW$, we followed \citet{2001A&A...379..976M} and considered 
\textsc{jstarmod}'s typical internal precisions (0.5-2 km\,s$^{-1}$ in velocity shifts and $\pm$ 5 km\,s$^{-1}$ in $v$~$\sin{i}$), 
the $rms$ in regions outside the chromospheric features (typically 0.01-0.03), and the standard deviations. The estimated 
errors for relatively strong emitters are in the range of 10-20\% but for low activity stars errors are larger. Taking into 
consideration that S/N is lower in the blue spectral region, errors in the chromospheric features at these wavelengths 
are larger.

\begin{figure}[h!]
\begin{center}
\includegraphics[width=9.cm, keepaspectratio]{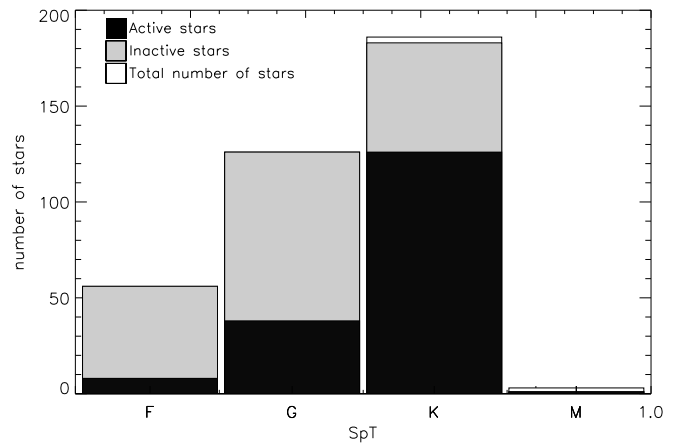}
\caption{Number of active (black) and inactive (grey) stars in the sample for each spectral type.}
\label{histograma}
\end{center}
\end{figure}

\begin{table*}
\caption{Inactive stars used as references in the subtraction technique to measure chromospheric activity}
\label{ref_activity}
\centering
\begin{tabular}{l l l l l l l l l l l l}
\hline\hline
\noalign{\smallskip}
\multicolumn{1}{c}{HIP}& SpT & \multicolumn{1}{c}{$B$--$V$} & $v$~$\sin{i}$ & Obs$^{\rm 1}$. & \multicolumn{1}{c}{$\log{\rm (F_{\rm H} + F_{\rm K})}$} &
\multicolumn{1}{c}{HIP}& SpT & \multicolumn{1}{c}{$B$--$V$} & $v$~$\sin{i}$ & Obs$^{\rm 1}$. & \multicolumn{1}{c}{$\log{\rm (F_{\rm H} + F_{\rm K})}$}\\
\noalign{\smallskip}
 & & & (km\,s$^{-1}$) & &\multicolumn{1}{c}{(erg~cm$^{\rm -2}$~s$^{\rm -1}$)} &&&& (km\,s$^{-1}$) &  & \multicolumn{1}{c}{(erg~cm$^{\rm -2}$~s$^{\rm -1}$)}\\
\noalign{\smallskip}
\hline
\noalign{\smallskip}
37279	&	F5V	&	0.420	&	5.38	&	Mc	&	0.76$^{\rm a}$	&	8102	&	G8V	&	0.730	&	8.00	&	Mc	&	0.15$^{\rm a}$, 0.14$^{\rm d}$	\\
910	&	F5V	&	0.489	&	4.88	&	S	&	...	&	79492	&	G8V	&	0.756	&	2.68	&	FO	&	0.12$^{\rm a}$, 0.11$^{\rm d}$	\\
78072	&	F6IV	&	0.480	&	11.42	&	FO	&	0.57$^{\rm a}$	&	95319	&	G8V	&	0.805	&	0.60	&	FO	&	-0.05$^{\rm a}$, 0.01$^{\rm d}$	\\
22449	&	F6V	&	0.475	&	19.22	&	Mc	&	0.71$^{\rm a,d}$	&	101997	&	G8V	&	0.730	&	3.50	&	S	&	0.16$^{\rm d}$	\\
40035	&	F7V	&	0.495	&	11.26	&	S	&	...	&	47080	&	G8V	&	0.779	&	6.97	&	FO/S/Mc	&	0.30$^{\rm a}$	\\
27072	&	F7V	&	0.498	&	...	&	FE	&	...	&	63366	&	G9V	&	0.780	&	...	&	FE	&	0.12$^{\rm a}$	\\
17147	&	F9V	&	0.538	&	9.70	&	FO	&	0.48$^{\rm d}$	&	74537	&	K0V	&	0.761	&	2.12	&	S	&	...	\\
57757	&	F9V	&	0.558	&	3.41	&	Mc	&	0.44$^{\rm d}$	&	40693	&	K0V	&	0.766	&	 $\leq$ 6.79	&	Mc	&	0.09$^{\rm d}$	\\
16852	&	F9V	&	0.568	&	2.67	&	Mc	&	0.39$^{\rm a}$, 0.37$^{\rm d}$	&	84195	&	K0	&	0.940	&	8.58	&	FO	&	-0.20$^{\rm d}$	\\
1599	&	F9V	&	0.576	&	$\leq$ 3.23	&	FE	&	0.45$^{\rm b}$	&	112190	&	K0	&	0.968	&	$\leq$ 3.35	&	FO	&	-0.12$^{\rm d}$	\\
64394	&	F9.5V	&	0.588	&	4.72	&	Mc	&	0.48$^{\rm a,d}$	&	3093	&	K0.5V	&	0.853	&	9.38	&	FO/S/Mc	&	-0.05$^{\rm a}$, -0.11$^{\rm d}$	\\
61317	&	G0V	&	0.589	&	2.00	&	Mc	&	0.43$^{\rm a}$, 0.41$^{\rm d}$	&	7981	&	K1V	&	0.834	&	6.50	&	FO/S/Mc	&	-0.01$^{\rm a}$, -0.02$^{\rm d}$	\\
77257	&	G0IV	&	0.596	&	3.00	&	Mc	&	0.37$^{\rm a,d}$	&	70016	&	K1V	&	0.867	&	8.15	&	S	&	-0.13$^{\rm a}$, -0.07$^{\rm d}$	\\
14632	&	G0V	&	0.606	&	3.15	&	Mc	&	0.33$^{\rm a}$, 0.32$^{\rm d}$	&	79190	&	K1V	&	0.843	&	$\leq$ 3.97	&	FE	&	...	\\
77801	&	G0V	&	0.624	&	...	&	FO/S	&	0.37$^{\rm d}$	&	71681	&	K1V	&	0.900	&	$\leq$ 3.52	&	FE	&	...	\\
1499	&	G0V	&	0.680	&	4.18	&	FO	&	0.21$^{\rm d}$	&	3765	&	K2V	&	0.885	&	6.71	&	FO/S/Mc	&	-0.04$^{\rm a}$	\\
67904	&	G0V	&	0.697	&	...	&	FE	&	0.27$^{\rm d}$	&	88972	&	K2V	&	0.886	&	4.82	&	S	&	-0.11$^{\rm a}$, -0.13$^{\rm d}$	\\
10644	&	G0.5V	&	0.603	&	3.93	&	Mc	&	0.52$^{\rm a}$, 0.50$^{\rm d}$	&	105152	&	K2V	&	1.028	&	3.69	&	FO	&	-0.33$^{\rm a}$	\\
29860	&	G0.5V	&	0.611	&	2.98	&	S	&	0.34$^{\rm a,d}$	&	114886	&	K2V	&	0.898	&	$\leq$ 3.18 	&	FO	&	-0.07$^{\rm d}$	\\
48113	&	G0.5IV	&	0.624	&	2.93	&	FO/S	&	0.23$^{\rm a}$, 0.27$^{\rm d}$	&	12114	&	K3V	&	0.918	&	6.45	&	S	&	-0.11$^{\rm a}$, -0.04$^{\rm d}$	\\
15371	&	G1V	&	0.600	&	$\leq$ 2.64	&	FE	&	0.42$^{\rm c}$	&	114622	&	K3/K4 V	&	1.000	&	2.10	&	FO/S	&	-0.29$^{\rm a}$	\\
53721	&	G1V	&	0.613	&	2.80	&	Mc	&	0.31$^{\rm a}$, 0.35$^{\rm d}$	&	78843	&	K3/K4V	&	1.059	&	6.24	&	S	&	-0.38$^{\rm d}$	\\
24813	&	G1.5IV	&	0.613	&	2.00	&	Mc	&	0.31$^{\rm a,d}$	&	73184	&	K4V	&	1.110	&	...	&	FE	&	-0.20$^{\rm d}$	\\
7918	&	G1.5V	&	0.620	&	2.10	&	FO/S/Mc	&	0.30$^{\rm a}$	&	113718	&	K4V	&	0.948	&	6.22	&	FO	&	...	\\
52369	&	G2/G3V	&	0.629	&	7.22	&	S	&	...	&	12929	&	K5	&	1.170	&	8.11	&	FO	&	0.01$^{\rm a}$, -0.07$^{\rm d}$	\\
79672	&	G2Va	&	0.648	&	$\leq$ 4.07	&	Mc	&	0.31$^{\rm a}$, 0.30$^{\rm d}$	&	50125	&	K5V	&	1.122	&	2.85	&	S	&	...	\\
27435	&	G4V	&	0.639	&	2.61	&	S	&	0.32$^{\rm d}$	&	54651	&	K5V	&	1.089	&	$\leq$ 0.53	&	S	&	...	\\
50505	&	G5	&	0.686	&	1.72	&	S	&	0.22$^{\rm d}$	&	83591	&	K5V	&	1.120	&	3.70	&	FO/S	&	-0.20$^{\rm d}$	\\
64924	&	G5V	&	0.709	&	4.09	&	S	&	0.16$^{\rm a}$, 0.15$^{\rm d}$	&	93871	&	K5V	&	1.050	&	4.07	&	FO	&	...	\\
171	&	G5V	&	0.660	&	3.00	&	Mc	&	0.30$^{\rm a}$	&	104214	&	K5V	&	1.160	&	4.72	&	FO/S	&	-0.22$^{\rm a}$	\\
5336	&	G5V	&	0.700	&	8.00	&	Mc	&	0.19$^{\rm a}$	&	80644	&	K6V	&	1.209	&	 $\leq$ 3.68	&	FO	&	-0.25$^{\rm a}$, -0.26$^{\rm d}$	\\
62523	&	G7V	&	0.706	&	10.60	&	S	&	0.42$^{\rm d}$	&	104217	&	K7V	&	1.360	&	1.70	&	FO/S	&	-0.50$^{\rm a}$	\\
2941	&	G7V	&	0.717	&	5.54	&	S	&	0.23$^{\rm a}$, 0.19$^{\rm d}$	&	54646	&	K8V	&	1.345	&	5.83	&	S	&	-0.37$^{\rm d}$	\\
58576	&	G8IV-V	&	0.757	&	...	&	Mc	&	0.03$^{\rm a}$, 0.11$^{\rm d}$	&	60661	&	M0V	&	1.451	&	...	&	FO	&	...	\\
14150	&	G8V	&	0.715	&	4.08	&	S	&	0.22$^{\rm d}$	&		&		&		&		&		&		\\
\noalign{\smallskip}
\hline
\end{tabular}
\begin{flushleft}
{\tiny
$^{\rm 1}$ Spectrograph used: Mc: Mc Donald; S: SARG; FO: FOCES; FE: FEROS \\
$^{\rm a}$ Duncan et al. 1991 \\
$^{\rm b}$ Henry et al. 1996 \\
$^{\rm c}$ Jenkins et al. 2006 \\
$^{\rm d}$ Wright et al. 2004}
\end{flushleft}
\end{table*}
In Fig. \ref{histograma}, we plot an histogram for the total number of stars of each spectral type and the
number that could be classified as active (displaying chromospheric features in the spectrum) or not active.
In Table \ref{tab:activity_ew}, we give the excess emission $EW$ and its error for the active stars in the sample. 
Of the complete sample of stars analysed (371 stars), 173 presented chromospheric activity 
features in their spectra. Of them, 8 were F type stars (14\% of the analysed F stars), 38 G type stars (30\% of the G type stars),
 126 K type stars (68\% of the analysed K stars) and 1 M star (33\% of the analysed M stars). 
From the remaining, 193 could be classified as inactive because of the lack 
of any chromospheric activity feature in their spectra. Of them, 48 are F type stars 
(86\% of the analysed F stars), 88 G type stars (70\% of the total G type stars), and 57 K stars (31\% of the analysed K type stars). 
The remaining 5 stars could not be classified as either active or inactive because of the low S/N of their spectra. 
We have included a column (Col. \# 8) in Table \ref{parameters} to specify whether the star can be considered as active or not active.
In 6 cases (5 K stars and 1 M star), we could not measure chromospheric activity due to the lack of a suitable 
non-active reference star to perform the subtraction technique, but chromospheric activity features were clearly present in the spectra. 
These stars were classified as active (and indicated with * in Table \ref{parameters}) but chromospheric activity was not measured. 
It is also important to mention that due to the configuration used with the SARG spectrograph
the spectral range corresponding to \ion{Ca}{ii} H \& K was not covered and these lines could not be measured.

We must comment on eight special cases: HIP 3093, HIP 3765, HIP 7981, HIP 54646, HIP 60866, 
HIP 62523, HIP 77408 and HIP 85810. These stars were observed more than once and with different instruments. Chromospheric activity 
features were detected in at least one of the observations, but not in all of them. Therefore, the stars are labelled as active and 
inactive depending on the observing run. In six of the cases (HIP 3093, HIP 3765, HIP 54646, HIP 60866, HIP 77408 and HIP 85810), 
the level of activity (when measured) is very low, which points to the use of a different reference star as the explanation for the
inability to detect emission features in the subtracted spectrum. Two of these stars (HIP 3093 and HIP 3765) show levels of chromospheric 
activity so low that are generally considered inactive and used as reference stars to subtract the photospheric contribution from the spectrum. 
In the remaining cases, HIP 7981 and HIP 62523, variability appears to be 
the cause. The star HIP 7981 was previously classified as variable by \citet{2007AJ....133..862H}. 
Both stars were observed in three observing runs, two of them closer in time than the other. The stars exhibit no
chromospheric emission features in observations carried out during the same epoch, whereas they do in data for  
other epoch. This indicates that the lack of features appears to be real and not attributable to a different choice of 
the reference star. Variability therefore presents itself as a plausible explanation of the lack of detected activity 
in two of the observations. These stars are marked with $\star$ in Table \ref{parameters}. 
Fluxes can be derived from the measured $EW$ by correcting the continuum flux
\begin{equation}
F_{\rm \lambda}=EW_{\rm \lambda}\,F_{\rm \lambda}^{\rm cont}\;\; \Longrightarrow\;\; \log F_{\rm \lambda} = \log (EW)\,+\,\log(F_{\rm \lambda}^{\rm cont}) ,
\end{equation}
where the continuum flux, $F_{\lambda}^{\rm cont}$, is obviously dependent on the wavelength and must therefore be determined for
 the region where the activity indicator line appears. We used the empirical relationships between $F_{\lambda}^{\rm cont}$ and 
colour index, $B$--$V$, \citep{1996PASP..108..313H} to compute $F_{\lambda}^{\rm cont}$ for each line and star. 
We note that the aforementioned relationships are linear for the spectral type  
range of the sample stars. In Table \ref{tab:activity_flux}, we give the absolute flux at the stellar surface and its error for 
the active stars in the sample.
\subsubsection{$R'_{\rm HK}$ index}
\label{s_Rhk}
\begin{figure}[!htbp]
\centering
\includegraphics[width=9.3cm, bb = 54 360 534 671,clip]{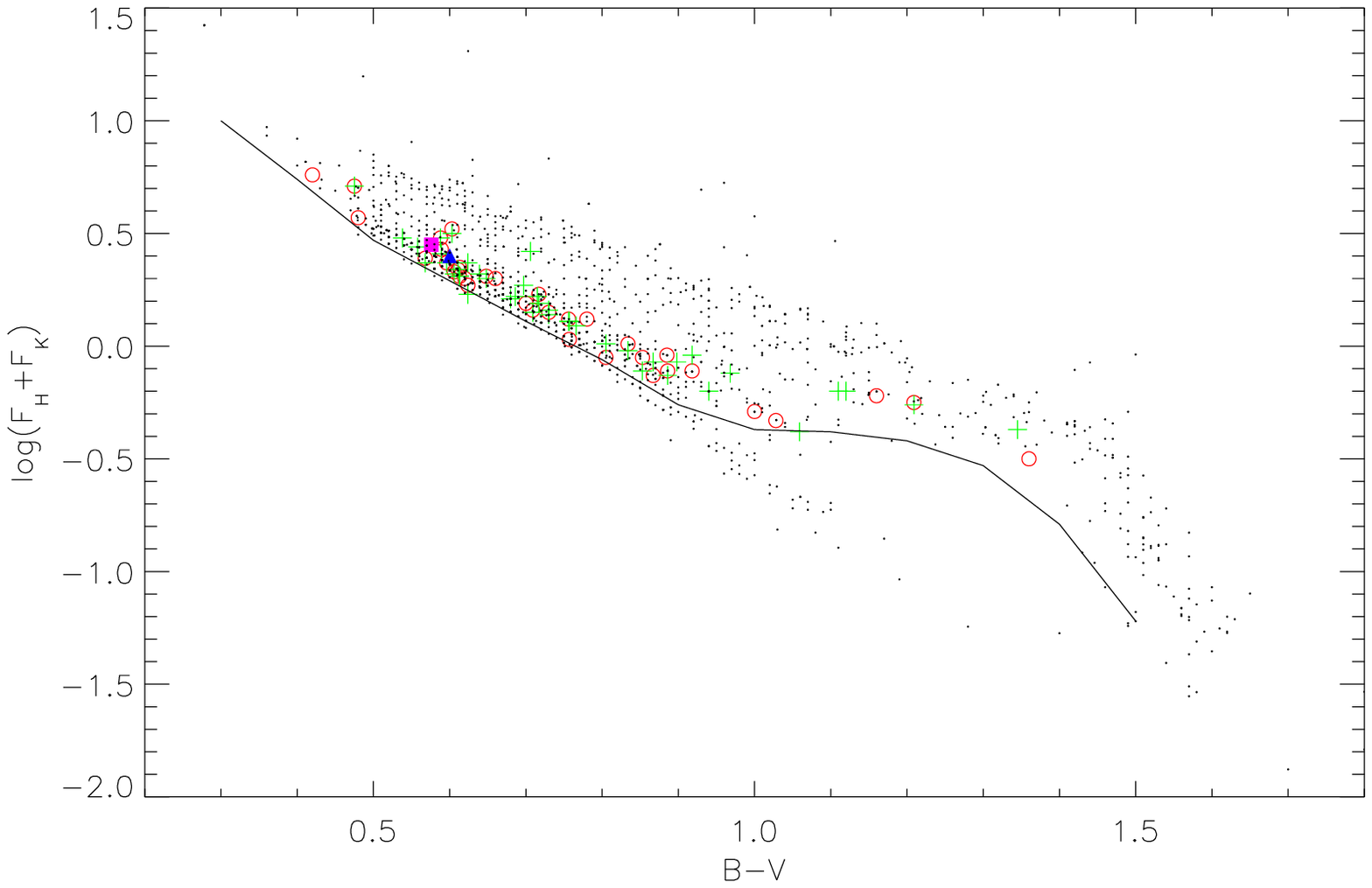}
\caption{\ion{Ca}{ii} H \& K surface flux $vs$ colour index $B$--$V$. Small dots represent \citet{1991ApJS...76..383D} 
and \citet{2004ApJS..152..261W} data. To represent the stars considered as reference in this work, we have used different symbols according 
to the source of F$_{\rm H}$+F$_{\rm K}$ values: circles for \citet{1991ApJS...76..383D} data, squares for \citet{1996AJ....111..439H} data, 
crosses for \citet{2004ApJS..152..261W} data, and triangles for \citet{2006MNRAS.372..163J} data. 
The curve represents the surface flux boundary obtained by \citet{1984A&A...130..353R}.} 
\label{minimun_surf_flux}
\end{figure}

Chromospheric activity has been traditionally studied using the $R'_{\rm HK}$ index, defined as the ratio of the emission from the chromosphere 
in the cores of the \ion{Ca}{ii} H \& K to the total bolometric emission of the star, where the prime denotes that subtraction of the 
photospheric contribution has been performed.It was first used by \citet{1984ApJ...279..763N}. 
This index was first measured using observed flux indices in the core of \ion{Ca}{ii} H \& K (corrected from the 
continuum signal) with the Mount Wilson H-K spectrophotometer \citep{1978PASP...90..267V}. 
These fluxes were corrected from the photospheric contribution using an empirical calibration 
with the colour index $B$--$V$ \citep{1984ApJ...279..763N}. Finally, the \ion{Ca}{ii} H and K line-core flux measurements 
were corrected from the minimum surface flux using a calibration with the colour index $B$--$V$ \citep{1984A&A...130..353R}, 
thus providing a measurement of the chromospheric contribution associated with magnetic activity. 
In principle, we could derive $R'_{\rm HK}$ directly from the measured fluxes in \ion{Ca}{ii} H \& K lines 
using the subtraction technique
\begin{equation}\label{RHK}
R'_{\rm HK} = \frac{F'_{\rm H} + F'_{\rm K}}{\sigma T_{\rm eff}^{\rm 4}}.
\end{equation} 
When the subtraction technique is applied, the residual chromospheric contribution of the reference star is also 
subtracted from the spectrum of the target star. The source of possible differences between the fluxes obtained using the subtraction 
technique and those obtained with the traditional method, is the difference in chromospheric emission between our reference stars 
and those used by \citet{1984A&A...130..353R} to compile his calibration. In Fig. \ref{minimun_surf_flux}, we have plotted the total surface 
flux in the \ion{Ca}{ii} H \& K lines for the reference stars used in this work, and the \citet{1984A&A...130..353R} calibration 
for main sequence stars. The values of total surface flux 
in the \ion{Ca}{ii} H \& K lines for the reference stars have been obtained from \citet{1991ApJS...76..383D}, \citet{1996AJ....111..439H}, 
\citet{2004ApJS..152..261W}, and \citet{2006MNRAS.372..163J} and are included in Table \ref{ref_activity}. We note that all our reference 
stars have \ion{Ca}{ii} H \& K fluxes close to the lower 
boundary defined by \citet{1984A&A...130..353R} adopted in subsequent studies. The maximum difference between the surface fluxes of 
the stars used as references in the present study and the \citet{1984A&A...130..353R} calibration is 0.2 dex, with the exception of K5 to 
K7 stars for which it is 0.35 dex. Since these differences account for differences of only 0.2 dex (0.35 dex for K5 to K7 stars) 
in R'$_{\rm HK}$, we can assume that the values of $R'_{\rm HK}$ obtained using the subtraction technique are comparable to those 
obtained with the traditional method.  

To determine the effective temperatures needed to convert total flux in \ion{Ca}{ii} H \& K into $R'_{\rm HK}$, we used 
the empirical calibrations with the colour index $B$--$V$ provided by \citet{2008oasp.book.....G}, which holds for the 
spectral type range of the target stars (0.00 $\leq$ $B$--$V$ $\leq$ 1.5).

%\tiny
\begin{table*}
\caption{Comparison of the classification of the stars as active or inactive with previous results.}
\label{concordancia_numeros}
\centering
\begin{tabular}{l c c c c c}
\hline\hline
%\noalign{\smallskip}
% & \multicolumn{2}{c}{\#} &  &  \multicolumn{2}{c}{\#}\\
\noalign{\smallskip}
\multicolumn{1}{c}{Dataset}& \multicolumn{2}{c}{\# of common stars} & & \multicolumn{2}{c}{\# of coincidences}\\
\cline{2-3}\cline{5-6}
\noalign{\smallskip}
 & \multicolumn{1}{c}{Active} & \multicolumn{1}{c}{Inactive}&  & \multicolumn{1}{c}{Active} & \multicolumn{1}{c}{Inactive}\\
\noalign{\smallskip}
\hline
\noalign{\smallskip}
\noalign{\smallskip}
Duncan et al. (1991)                & 52 & 90 & & 40 & 75 \\
%\multicolumn{1}{c}{(1991)}    &    &    & &    &    \\
Strassmeier et al. (2000)           & 37 & 21 & & 31 & 20 \\
%\multicolumn{1}{c}{(2000)}    &    &    & &    &    \\
Wright et al. (2004)                & 34 & 92 & & 30 & 84 \\
%\multicolumn{1}{c}{(2004)}    &    &    & &    &    \\
Hall et al. (2007)                  & 21 & 35 & & 14 & 31 \\
%\multicolumn{1}{c}{(2007)}    &    &    & &    &    \\
Mamajek \& Hillenbrand (2008)       & 27 & 54 & & 23 & 47 \\
%\multicolumn{1}{c}{(2008)}    &    &    & &    &    \\
\hline
\end{tabular}
\end{table*}

\begin{figure}[h!]
\centering
\includegraphics[width=9.0cm,bb=35 360 281 610,clip]{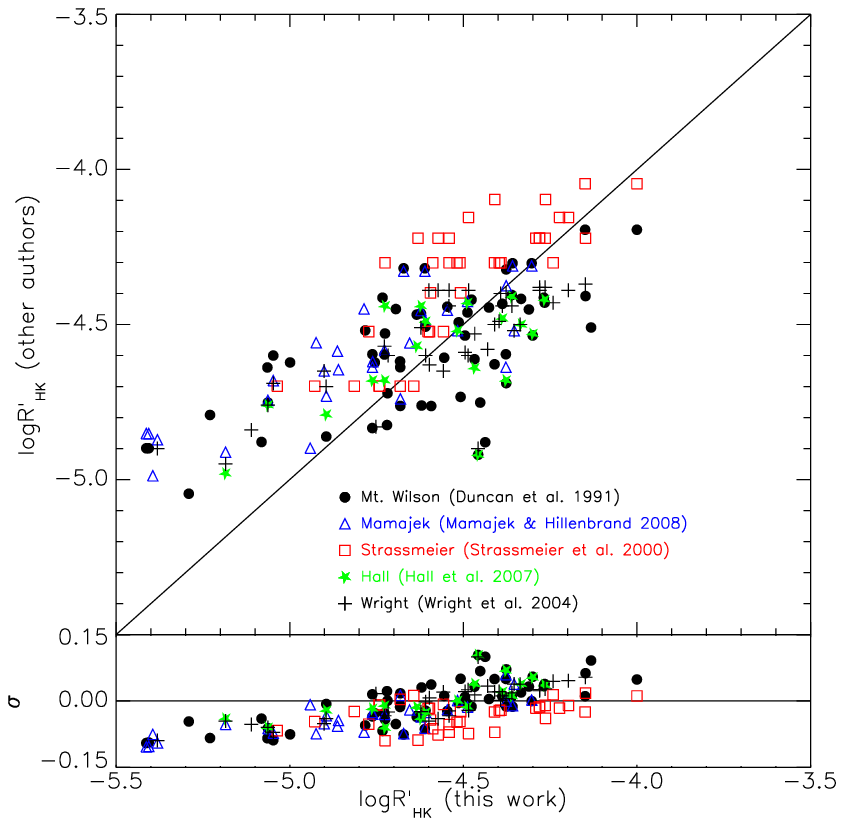}
\caption{Comparison of $R'_{\rm HK}$ index obtained in this paper and that obtained with the Mount Wilson H-K spectrophotometer
\citep{1978PASP...90..267V} and similar techniques. Different symbols are used for \citet{1991ApJS...76..383D}, \citet{2000A&AS..142..275S}, 
\citet{2004ApJS..152..261W}, \citet{2007AJ....133..862H}, and \citet{2008ApJ...687.1264M} data. In the lower panel, we have 
plotted $\sigma$ (as described in the text).}
\label{comparison}
\end{figure}
\subsection{Comparison with previous results}
\label{subsection:comparison}
To test whether the transformation is consistent with those values of $R'_{\rm HK}$ computed using photometry (or a technique to mimic 
photometric results using spectroscopic data), we compared our data to those obtained by \citet{1991ApJS...76..383D}, 
\citet{2000A&AS..142..275S}, \citet{2004ApJS..152..261W}, \citet{2007AJ....133..862H}, and \citet{2008ApJ...687.1264M}. 
The comparison is plotted in Fig. \ref{comparison}, where 
\begin{equation}
\sigma = \frac{(\log R'_{\rm HK} (\rm other\;authors) - \log R'_{\rm HK} (\rm this\;work))}{\log R'_{\rm HK} (\rm this\;work)}.
\end{equation}
The dispersion observed in Fig. \ref{comparison} is compatible with variations in activity levels with time. To determine if there are 
systematic differences between our data and any of the five data sets analysed, we have plotted $\sigma$ in  Fig. \ref{comparison}. 
The closer the value of $\sigma$ to 0, the smaller the difference between our values and those from other authors. In addition, we 
performed a Kolmogorov-Smirnov test to determine whether our data and those obtained by other authors are equivalent. 
The values of the estatistical estimator $D$ obtained when applying the Kolmogorov-Smirnov test with Mt. Wilson 
($n$ = 59), Strassmeier ($n$ = 37), Wright ($n$ = 38), Hall ($n$ = 19) and Mamajek ($n$ = 31) 
data are 0.152, 0.351, 0.236, 0.263 and 0.322, respectively. These results indicate that the null hypothesis, $i$. $e$., both samples 
are equivalent, could not be rejected at a level of significance less than 65 \% in the Mt. Wilson, Wright, and Hall cases. 
This clearly means that $\log R'_{\rm HK}$ values obtained using the traditional method are equivalent to those obtained in this study. 
As might be expected, discrepancies are larger when we compare our results with those presented in \citet{2008ApJ...687.1264M}, because the 
latter constitutes a compilation of values obtained from different sources. Differences between our results and 
\citet{2000A&AS..142..275S} are also larger than the rest. Again this is not surprising, taking into account that the  
method used to correct from the photospheric contribution is different than in the rest of the cases. The \citet{2000A&AS..142..275S} 
values therefore do not necessarily reproduce the original $\log{R'_{\rm HK}}$ values. 
It is also important to note that discrepancies are larger for less active stars, in particular for stars with measured 
$\log R'_{\rm HK} \le -4.9$. This result is again expectated considering that for stars
with low activity levels, errors in the photospheric contribution correction become more apparent. 

%\tiny
\begin{table*}
\caption{Stars for which our classification as active or inactive differs from that of other authors.}
\label{concordancia}
\centering
\begin{tabular}{l l l l c c l c l }
\hline\hline
\noalign{\smallskip}
\multicolumn{1}{c}{HIP}& \multicolumn{1}{c}{HD} & \multicolumn{2}{c}{This work} & & \multicolumn{4}{c}{Previous results}\\
\cline{3-4} \cline{6-9}\\
 & & \multicolumn{1}{c}{Activity} & $\log R'_{\rm HK}$ & & Activity & $\log R'_{\rm HK}$ & Activity & \multicolumn{1}{c}{$\log R'_{\rm HK}$}\\
\noalign{\smallskip}
\hline
%\endhead
\hline
\noalign{\smallskip}
3765    &  4628    & active   & -5.40 & & inactive & -4.89$^{1,5}$ & ... & ... \\ 
5286    &  6660    & active   & -4.57 & & inactive & -4.76$^{1}$ & ... & ... \\
7751    &  10360   & active   & -4.94 & & inactive & -4.90$^{5}$ & ... & ... \\
7981    &  10476   & active   & -5.19 & & inactive & -4.95$^{3}$,-4.98$^{4}$,-4.91$^{5}$ & ... & ... \\ 
10644   &  13974   & inactive &  ...  & & ...      & ...         & active & -4.64$^{1,5}$, -4.71$^{3}$, -4.69$^{4}$\\
12929   &  17230   & inactive &  ...  & & ...      & ...         & active & -4.55$^{1}$\\
15442   &  20619   & active   & -4.75 & & inactive & -4.83$^{3}$ & ... & ...\\
15457   &  20630   & inactive &  ...  & & ...      & ...         & active & -4.41$^{1,5}$, -4.71$^{3}$, -4.40$^{4}$\\
17420   &  23356   & inactive &  ...  & & ...      & ...         & active & -4.69$^{2}$\\
19422   &  25665   & inactive &  ...  & & inactive & -4.86$^{1}$ & active & -4.69$^{2}$\\
19849   &  26965   & active   & -5.38 & & inactive & -4.87$^{5}$, -4.90$^{3}$ &... & ...\\
20917   &  28343   & active   & -4.62 & & inactive & -4.76$^{1}$ & ... & ... \\
22449   &  30652   & inactive &  ...  & & inactive & -4.79$^{1}$ & active & -4.65$^{3}$\\
23311   &  32147   & active   & -5.29 & & inactive & -5.75$^{1}$ & ... & ... \\
36551   &  59582   & active   & -4.44 & & inactive & -4.88$^{1}$ & ... & ... \\
40693   &  69830   & active*  &  ...  & & inactive & -4.95$^{3,5}$ & ... & ... \\ 
42173   &  72946   & inactive &  ...  & & ... &  ... & active & -4.46$^{1}$ \\ 
41484   &  71148   & inactive &  ...  & & inactive & -4.95$^{3}$, -4.94$^{4}$  & active &     -3.65$^{1}$\\
43726   &  76151   & inactive &  ...  & & ... & ...  & active &      -4.59$^{1}$, -4.66$^{4}$\\
46509   &  81997   & inactive &  ...  & & ... & ...  & active &      -4.67$^{4}$\\
46853   &  82443   & inactive &  ...  & & ... & ...  & active &  -4.01$^{1}$, -4.05$^{2}$ \\ 
49699   &  87883   & inactive &  ...  & & ... & ...  & active &      -5.00$^{2}$\\
56452   &  100623  & active*  &  ...  & & inactive & -4.89$^{3,5}$ & ... & ... \\ 
56997   &  101501  & inactive &  ...  & & ... & ...  & active &      -4.54$^{1,5}$, -4.55$^{3}$, -4.62$^{4}$\\
64394   &  114710  & active   & -5.06 & & inactive & -4.75$^{1}$, -4.76$^{3,4}$ & active & -4.74$^{5}$\\
64792   &  115383  & inactive &  ...  & & ... & ... & active &       -4.45$^{1}$, -4.40$^{3}$, -4.47$^{4}$\\
67275   &  120136  & active   & -4.89 & & inactive & -4.86$^{1}$, -4.79$^{3}$ & active & -4.73$^{5}$\\
68337   &  122120  & active   & -4.68 & & inactive & -4.82$^{1}$ & ... & ...\\
69701   &  124850  & inactive &  ...  & & inactive & -4.75$^{1}$ & active &      -4.69$^{4}$\\
72875   &  131582  & active   & -4.45 & & inactive & -4.75$^{1}$ & ... & ...\\
73695   &  133640  & inactive &  ...  & & ... & ... & active & -4.62$^{1}$, -4.64$^{5}$ \\
81375   &  149806  & inactive &  ...  & & inactive  & -4.83$^{3}$ & active & -4.70$^{2}$ \\ 
84195   &  155712  & inactive &  ...  & & ... & ... & active &       -4.69$^{2}$\\
85810   &  159222  & active   & -4.48 & & inactive & -4.92$^{1}$, -4.90$^{3}$& ... & ...\\
86400   &  160346  & active   & -4.76 & & inactive & -4.83$^{1}$ & ... & ... \\ 
88622   &  165401  & inactive &  ...  & & ... & ... & active &       -4.61$^{1}$\\
96285   &  184489  & active   & -5.08 & & inactive & -4.88$^{1}$ & ... & ...\\
97649   &  187642  & inactive &  ...  & & ... & ... & active & -4.47$^{1}$ \\ 
99461   &  191408  & active   & -5.39 & & inactive & -4.99$^{5}$ & ... & ...\\
101955  &  196795  & inactive &  ...  & & inactive & -4.78$^{1}$ & active & -5.00$^{2}$\\
104092  &  200779  & active   & -5.14 & & inactive & -5.14$^{1}$ & ... & ...\\
108156  &  208313  & active   & -4.68 & & inactive & -4.76$^{1}$ & ... & ...\\
114886  &  219538  & active   & -4.84 & & inactive & -4.84$^{3}$ & ... & ...\\
\noalign{\smallskip}
\hline
\multicolumn{4}{l}{\tiny $^1$ Duncan et al. (1991)}\\
\multicolumn{4}{l}{\tiny $^2$ Strassmeier et al. (2000)}\\
\multicolumn{4}{l}{\tiny $^3$ Wright et al. (2004)}\\
\multicolumn{4}{l}{\tiny $^4$ Hall et al. (2007)}\\
\multicolumn{4}{l}{\tiny $^5$ Mamajek \& Hillenbrand (2008)}\\
\multicolumn{9}{l}{\tiny * Activity features in the spectrum but values not measured due to the lack of a suitable reference star.}
\end{tabular}
\end{table*}
%\normalsize
%
We also compared our results with those obtained by the mentioned authors \citep{1991ApJS...76..383D,2000A&AS..142..275S,2004ApJS..152..261W,
2007AJ....133..862H, 2008ApJ...687.1264M}. In Table \ref{concordancia_numeros}, we summarise the number of active and inactive targets that 
we share with the mentioned authors as well as the number for which our classification is convergent. As mentioned in Sect. \ref{s_ew_flux} 
we classified a star as active when chromospheric features were present in the spectrum. To classify the stars observed by other authors, 
we used \citet{1999ApJ...524..295S} criterion to differentiate between active ($\log R'_{\rm HK}$ $>$ - 4.75) and inactive 
($\log{R'_{\rm HK}}$ $\leq$ -4.75) stars with the exception of \citet{2000A&AS..142..275S} data, for which the author provides his own classification. 
For \citet{1991ApJS...76..383D}, we obtained similar results for 84\% of the inactive and  77\% active stars. We obtained similar results to 
\citet{2000A&AS..142..275S} data for 95\% of the inactive stars and 84\% of the active ones. 
Concerning \citet{2004ApJS..152..261W} and based on the \citet{1999ApJ...524..295S} criterion, we reached agreement for 88\% of the active stars 
and 91\% of the inactive ones. For the \citet{2007AJ....133..862H} data, we obtained similar results for 89\% of the inactive stars and 
67\% of the active ones. Finally, when comparing our results to those of \citet{2008ApJ...687.1264M}, we reached an agreement for 87\% of the 
common inactive stars, and for 85\% of the actives ones. 

In Table \ref{concordancia}, we give details of the values of $\log{R'_{\rm HK}}$ found for those stars for which our classification as active 
or inactive differs from that of \citet{1991ApJS...76..383D}, \citet{2000A&AS..142..275S}, \citet{2004ApJS..152..261W}, \citet{2007AJ....133..862H}, 
or \citet{2008ApJ...687.1264M}. We note that with the exception of HIP 41484, which was classified as a {\itshape high-activity variable} 
by \citet{2007AJ....133..862H}, in all cases the measured values correspond to the low activity domain ($\log{R'_{\rm HK}}$ $\leq$ -4.40). 
It is important to mention that we 
classify a star as active or inactive after inspecting the spectrum from which the photospheric contribution has been subtracted. This means 
that every star showing chromospheric activity features will be considered active, regardless of the weakness of the activity 
levels that we measure. On the other hand, \citet{1999ApJ...524..295S} criterion
is based on the value obtained after measuring the activity. This implies that some of the stars that we have considered as active (because they show emission
features in the spectrum) should be reclassified as inactive after applying the aforementioned criterion. We prefer to maintain our criterion, and
consider inactive only those stars that did not exhibit emission features in the subtracted spectrum. Taking this into account, we can consider
the agreement between data obtained in the present work and that previously reported as fairly good.
\subsection{Spectral types}
\label{s_spectral_types}
As mentioned in Sect. \ref{s_ew_flux}, when performing the subtraction technique to measure chromospheric activity the use of a 
reference non-active star with similar physical properties (temperature and surface gravity) to those of the target star is necessary. 
Therefore, we were able to determine the spectral types of the target stars by comparing their spectra with those of the reference 
stars. The process began by assuming a spectral type for each star and applying the subtraction technique using as a reference an inactive 
star of similar spectral type and luminosity class. We then compared the non-chromospheric lines of the original and morphed spectra 
to test whether the spectral types of the star and that of the reference were really the same. In this way, we could correct the assumed spectral 
type of each star. We estimate the errors to be of one spectral subtype. 

In theory, if both stars have the same spectral type, the resultant (subtracted) spectrum should be null. 
In reality, the subtracted spectrum exhibit some noise, because of the small differences in metallicity and/or gravity
and when the S/N of one of the spectra is low. Nevertheless, small differences in metallicity (only population I stars 
were observed) and gravity (the cutoff of $\pm$ 1 mag from the Main Sequence (see Sect. \ref{s_sample}) corresponds to variations 
of $\pm$ 0.2 in $\log{g}$) are lower than those produced by the difference of one spectral subtype, which is the estimated 
error in the spectral type determination. Our results are shown in Table \ref{parameters}.
\section{Discussion}
\subsection{Flux-flux relationships}
\label{flux-flux}
Although chromopsheric activity has been traditionally studied using the $R'_{\rm HK}$ index, we have already pointed out that longer 
wavelengths provide noteworthy advantages when exoplanet searches are to be performed. The impact of chromospheric active regions 
on radial velocity variations appear to be smaller when the red region of the spectrum is considered \citep{2007A&A...473..983D}.
Moreover, the S/N in the red region of the spectrum is higher for cool stars. We have measured activity 
levels using activity tracers throughout the optical spectrum, including the IRT \ion{Ca}{ii} lines. 

After measuring chromospheric activity in different indicator lines, we have analysed the relationships between their fluxes. This 
approach was first introduced to study the magnetic structure of cool stars \citep{1987A&A...172..111S,1991A&A...252..203R} 
by comparing fluxes in chromospheric and coronal indicator lines. Subsequent studies generalised the method and analysed 
the relationship among different chromospheric indicators. The most widely studied relationship is that between H$_{\alpha}$ core emission 
and the total surface flux in \ion{Ca}{ii} H \& K lines \citep{1990ApJS...72..191S,1990ApJS...74..891R,2007A&A...469..309C,2009AJ....137.3297W}. 
Several studies have obtained fluxes in other chromospheric indicators lines, such as the \ion{Ca}{ii} infrared triplet, for binary 
\citep{1995A&A...294..165M,1996A&A...312..221M, 1996ASPC..109..657M} and single \citep{1993MNRAS.262....1T,2005ESASP.560..775L,
2007A&A...466.1089B} stars. The aforementioned studies were either centred on a specific spectral type range \citep{1993MNRAS.262....1T,
2009AJ....137.3297W} or analysed the relation between total fluxes, instead of that between each of the indicator lines.

\begin{figure*}[!htbp]
\centering
\includegraphics[width=9cm, keepaspectratio]{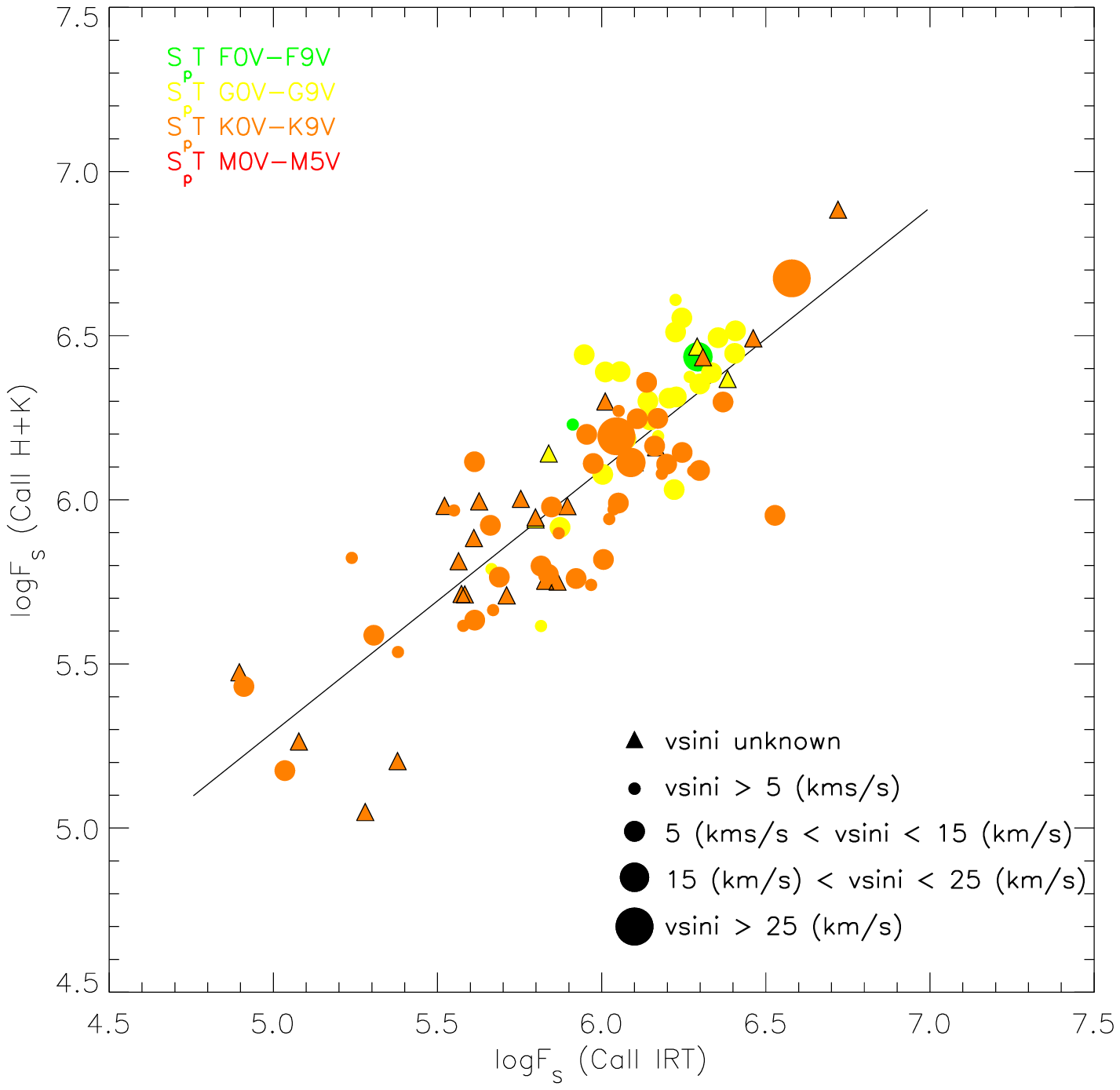}
\hfill
\includegraphics[width=9cm, keepaspectratio]{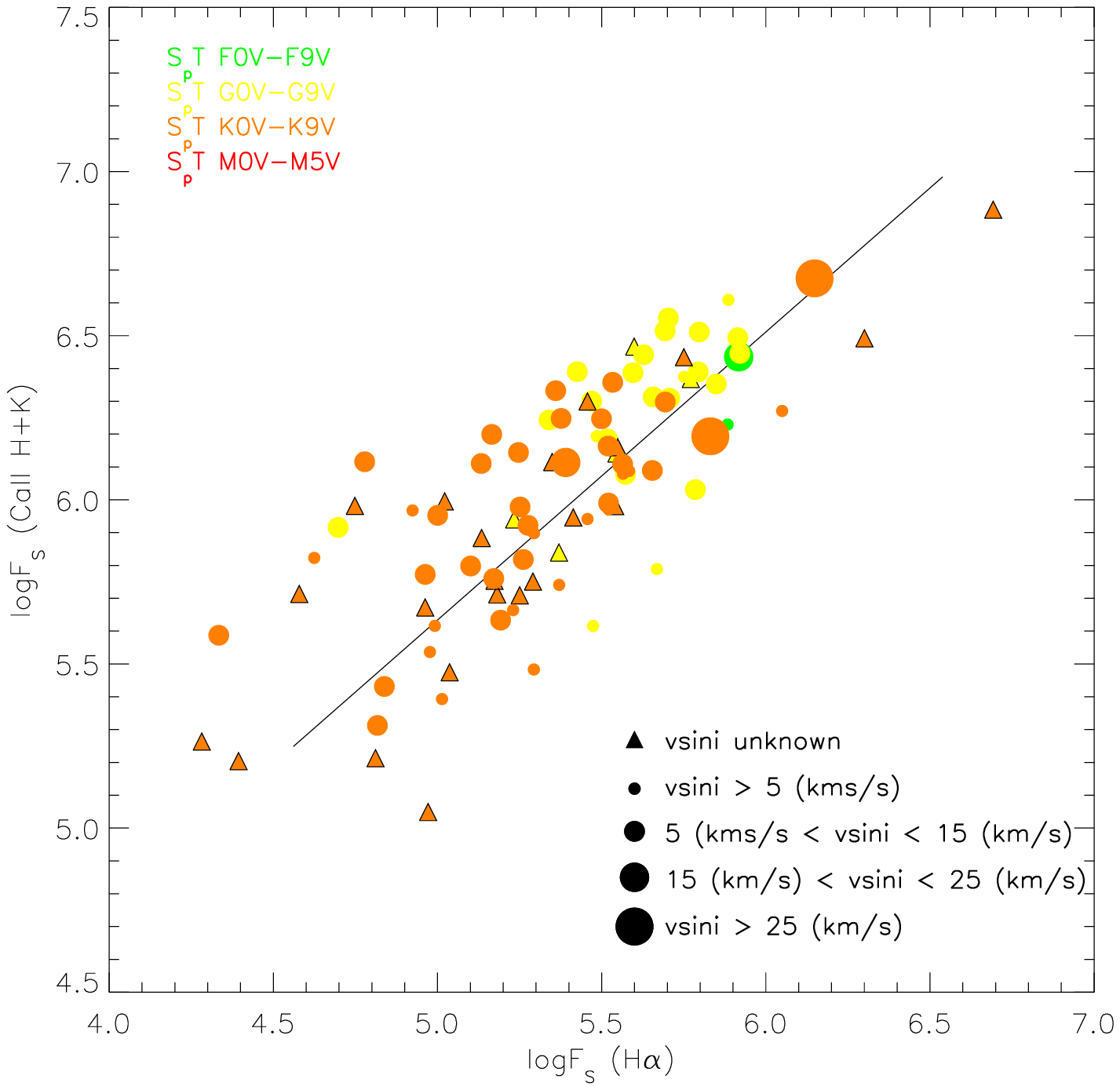}
\caption{Flux-flux relationship between the total flux in H\&K \ion{Ca}{ii} and \ion{Ca}{ii} IRT ({\bfseries left})
and the total flux in H\&K \ion{Ca}{ii} and \ion{Ca}{ii} IRT ({\bfseries right}). Symbol sizes
increase with increasing rotational velocity (triangles are used when $v$~$\sin{i}$ could not be determined).
Colors are used to discern different spectral types.}
\label{flux}
\end{figure*}

We have obtained empirical power-law relations between pairs 
of chromospheric indicator lines by fitting the data shown in Figs. \ref{flux} and Table \ref{flux_all} to an equation
\begin{equation}
\log F_{\rm 1} = c_{\rm 1} + c_{\rm 2} \log F_{\rm 2},
\end{equation}
where F$_{\rm 1}$ and F$_{\rm 2}$ are the fluxes of two different lines. We present the coefficients and the correlation coefficient 
($R$) of such relationships in Table \ref{tab:flux_rel}. 

In this context, the present study represents a significant extension in terms of spectral type range and number 
of stars. Moreover, we present relationships between each pair of chromospheric 
indicator lines in the optical range. These relationships have an enormous potential given that they 
permit the transformation between any pair of chromospheric activity indicator lines, in particular the transformation of 
\ion{Ca}{ii} H \& K fluxes to other more convenient ones. They might be extremely useful 
when using traditional photometric activity data, $i$.$e$., similar to that obtained by \citet{1978PASP...90..267V}, or when using 
spectroscopic data in which not all the chromospheric features are present. We have used them to obtain $\log R'_{\rm HK}$ when 
\ion{Ca}{ii} H or K lines could not be measured (see Sect. \ref{s_jitter}). 

\begin{table*}
\caption{Linear fit coefficients for each flux-flux relationship}
\label{tab:flux_rel}
\centering
\begin{tabular}{l l c c c}
\hline\hline
\noalign{\smallskip}
$\log F_{\rm 1}$ & $\log F_{\rm 2}$ & c$_{\rm 1}$ & c$_{\rm 2}$ & R\\
\hline
\noalign{\smallskip}
 \ion{Ca}{ii} H & \ion{Ca}{ii} K & -0.16 $\pm$ 0.20 & 1.01 $\pm$ 0.03 & 0.897\\
 \ion{Ca}{ii} H & H$\alpha$ & 1.95 $\pm$ 0.29 & 0.69 $\pm$ 0.05 & 0.736\\
 H$\alpha$  & \ion{Ca}{ii} K & -0.14 $\pm$ 0.44 & 0.95 $\pm$ 0.08 & 0.775\\
 \ion{Ca}{ii}IRT {\tiny ($\lambda$8498\AA)} & \ion{Ca}{ii} IRT {\tiny ($\lambda$8542 \AA)}& -0.15 $\pm$ 0.16 & 1.01 $\pm$ 0.03 & 0.894\\
 H$\alpha$  & \ion{Ca}{ii} IRT {\tiny ($\lambda$8542 \AA)} & -0.06 $\pm$ 0.29 & 0.98 $\pm$ 0.05 & 0.818\\
 \ion{Ca}{ii} H & \ion{Ca}{ii}IRT  {\tiny ($\lambda$8542 \AA)} & 1.27 $\pm$ 0.30 & 0.80 $\pm$ 0.05 & 0.830\\
 \ion{Ca}{ii} (H + K) & \ion{Ca}{ii} IRT ({\tiny $\lambda$8489 \AA + $\lambda$8542 \AA +$\lambda$8662 \AA})
& 1.30 $\pm$ 0.32 & 0.80 $\pm$ 0.05 & 0.852\\
 \ion{Ca}{ii} (H + K) & H$\alpha$ & 2.37 $\pm$ 0.30 & 0.68 $\pm$ 0.06 & 0.748\\
\noalign{\smallskip}
\hline
\end{tabular}
\end{table*}

%\onecolumn
%\twocolumn

\subsection{Predicted radial velocity jitter}
\label{s_jitter}

The most fruitful technique for detecting extrasolar planets has been the radial velocity method. As instrumental improvements and 
technique refinements have improved precisions in the m\,s$^{-1}$ domain, the analysis and minimization of the impact of $RV$ noise sources has become 
more important. There are two different kinds of $RV$ perturbations: the random and systematic measurement effects, and the intrinsic stellar 
variations. The former can be reduced by improving spectrographs and performing robust statistical analysis. The latter, however, includes 
several phenomena \citep{2009AIPC.1094..152S} and must be handled carefully. The $RV$ noise sources can lead to a false planet detection (if they 
produce a periodic signal over a few orbital periods) or prevent planet detection (if the perturbation is larger than the orbital $RV$ 
variation). Following \citet{2005ApJ...620.1002N}, the minimum detectable exoplanet mass with the $RV$ method is
\begin{equation}\label{minimun_mass}
M_{\rm min} \propto M_pN^{\rm -1/2}_{\rm obs}(\sigma^{\rm 2}_i+\sigma^{\rm 2}_{\rm rv})^{\rm 1/2}P^{\rm 1/3}M^{\rm 2/3}_{\rm *},
\end{equation}
where $M_{\rm p}$ is the planet mass, $M_{\rm *}$ is the stellar mass, $N$ is the number of observations, $P$ is the orbital period, and $\sigma_{\rm i}$ and 
$\sigma_{\rm rv}$ are the $rms$ instrumental error and $rms$ velocity jitter caused by stellar sources, respectively. Therefore, for a given system, 
the minimum detectable mass will be limited by the number of observations and the dominant noise source, $\sigma_{\rm i}$ or $\sigma_{\rm rv}$. 
Since spectrographs have been improved to achieve $\sigma_{\rm i}$ $\sim$ 1 m\,s$^{-1}$, the true limiting factor in Eq. \ref{minimun_mass} 
is the intrinsic stellar noise.

Stellar $RV$ variations are produced by different magnetic-activity-related phenomena: convection \citep{2009AIPC.1094..152S}, 
starspots \citep{1997ApJ...485..319S}, magnetic plage/network \citep{2003ASPC..294...65S} and flares \citep{2009AIPC.1094..152S,
2009A&A...498..853R}. Although bisector analysis may sometimes lead to the confirmation of a planet orbiting a star 
even when $RV$ jitter is present \citep{2006A&A...449..417S,2007ApJ...660L.145S,2008Natur.451...38S}, the latter technique is not always successful 
\citep{2008A&A...489L...9H,2009arXiv0912.2643F}. Several authors have studied the impact of activity on $RV$ jitter using $R'_{\rm HK}$ as a $proxy$ 
\citep{1998ApJ...498L.153S,2000A&A...361..265S,2002AJ....124..572P,2003csss...12..694S,2005PASP..117..657W,2006PASP..118..706P,2009arXiv0912.2901S}.
In particular, \citet{1998ApJ...498L.153S} and \citet{2000A&A...361..265S} compiled empirical relationships between $\sigma_{\rm rv}$ 
and $R'_{\rm HK}$ for stars in the Lick $v_r$ survey \citep{1998ARA&A..36...57M} and the Geneva extrasolar planet search programme. 
We used these relationships to obtain the expectable $RV$ jitter for the active stars in the sample. Results are given in Table 
\ref{jitter}. We note that, while \citet{2000A&A...361..265S} obtained individual relationships for G and K stars, 
\citet{1998ApJ...498L.153S} found that both type of stars exhibited similar trend. Agreement between the values obtained 
using each method is therefore fairly good except for K stars, which present larger $\sigma_{\rm rv}$ values for \citet{1998ApJ...498L.153S} 
than for \citet{2000A&A...361..265S} relationships.

It is important to mention that both relationships were obtained using stars with moderate activity levels (-5.0 $\leq$ 
$\log R'_{\rm HK}$ $\leq$ -4.0). We assumed that the linear fit holds for more active stars and applied the relations to 
five stars that have $\log R'_{\rm HK}$ $>$ -4.0. Whether this is valid or not is an open question that must be studied by 
recalibrating these relations by including a wider variety of activity levels. In our study, we considered chromospheric 
activity measurements in spectral ranges that contain some advantages for cool stars, $i.e.$ the \ion{Ca}{ii} IRT lines. 
New calibrations with different indices would be very beneficial to the community. In the context of this aim, our sample represents 
a large and varied (in terms of activity levels and activity indicators) set of stars but due to the unavailability of $\sigma_{rv}$ data 
(not public) we could not perform this analysis.

By applying the aforementioned empirical relationships and the derived values of $R'_{\rm HK}$, we calculated the expected $RV$ jitter for 
the stars in the sample, which we present in Table \ref{jitter}. For stars for 
which both \ion{Ca}{ii} H and \ion{Ca}{ii} K could be measured, $R'_{\rm HK}$ could be directly derived as described in Sect. \ref{s_Rhk}. 
However, in some cases, measurement of one of the \ion{Ca}{ii} lines (or both) was not possible 
due to low S/N or the presence of cosmic rays. In these cases, we used the empirical relationships obtained in Sect. \ref{flux-flux} to 
transform the total flux in \ion{Ca}{ii} IRT into \ion{Ca}{ii} (H + K) flux. We chose to use the \ion{Ca}{ii} IRT index because the dispersion 
between both indices is clearly lower than in the relation between the \ion{Ca}{ii} (H + K) and H$\alpha$ fluxes. However, when one or more 
lines in the triplet could not be measured, we used the flux in H$\alpha$ to infer that in \ion{Ca}{ii} (H + K). It is important to mention 
that the stars observed with the SARG spectrograph constitute a special case. With the configuration of the spectrograph we used, the spectral range 
containing \ion{Ca}{ii} H \& K is not available in the spectrum. The photospheric contribution correction for the orders containing 
the \ion{Ca}{ii} IRT lines was also less accurate than in the H$\alpha$ order. Consequently, we chose to use the empirical relationship between 
\ion{Ca}{ii} (H + K) and H$\alpha$ for these stars. The $R'_{\rm HK}$ values obtained are given in Table \ref{tab:activity_flux}. 
As mentioned in Sect. \ref{subsection:comparison}, discrepancies between our derived values of $\log{R'_{\rm HK}}$ and 
those obtained using a traditional method become important for stars with low activity levels. Given that the \citet{1998ApJ...498L.153S} 
and \citet{2000A&A...361..265S} relationships were obtained using traditional $\log{R'_{\rm HK}}$ data, the results obtained when applying 
them to the least active stars should be interpreted carefully. We have marked all stars with $\log{R'_{\rm HK}} \le -4.9$ in Table 
\ref{tab:activity_flux} with the symbol \ddag.

We cross-correlated our sample with the exoplanet database\footnote{\tt http://exoplanet.eu} and found that out of the total sample 
of 371 stars, 17 have confirmed exoplanets orbiting around them. As expected, all of them are either inactive stars, $i.e.$, 
HIP 1499 (HD 1461 b), HIP 7513 ($\upsilon$ And b), HIP 43587 (55 Cnc b), HIP 49699 (HD 87833 b), HIP 53721 (47 UMa b), HIP 64924 (61 Vir b), 
HIP 65721 (70 Vir b), HIP 109378 (HD 210277 b) and HIP 116727 ($\gamma$ Cephei b); or have very low activity levels, $i.e.$, HIP 3093 (HD 3651 b), 
HIP 10138 (Gl 86 b), HIP 16537 ($\epsilon$ Eri b), HIP 71395 (HD 128331 b), HIP 80337 (HD 147513 b), HIP 99711 (HD 192263 b) and HIP 113357 
(51 Peg b). For HIP 40693 (HD 69830 b), chromospheric activity could not be measured because of the lack of 
an inactive reference star to apply the subtraction technique, but chromospheric features were visible in the spectrum. 
The stars with known extrasolar planets are marked with \dag\, in Table \ref{parameters}.

\subsection{Applicability to transit searches}

Transit searches for exoplanets are also affected by the presence of active regions on the surface of a star 
\citep{1997ApJ...474..503H,1997ApJ...474L.119B,2000ApJ...531..415H}. \citet{2004A&A...414.1139A} used a Sun-based model 
to predict the ``stellar background'' using chromospheric activity (as given by $R'_{\rm HK}$). 
In the solar case, the noise spectrum for chromospheric
irradiance variations at frequencies lower than $\sim$ 8 mHz, commonly referred to as 
``solar background'', is frequently modelled by a sum of power laws, in which the number of terms, $N$, varies from one to five
depending on the frequency coverage \citep{1994SoPh..152..247A}, $i.e.$, 
\begin{equation}
P(\nu) = \sum_{i=1}^{\rm N} P_{\rm }i = \sum_{i=1}^{\rm N} \frac{A_{\rm i}}{1+(B_{\rm i}\nu)^{C_{\rm i}}},
\end{equation}
where $\nu$ is frequency, $A_{\rm i}$ is the amplitude of the $i$th component, $B_i$ is its characteristic timescale, and $C_{\rm i}$ is the slope
of the power law. For a given component, the power remains approximately constant on timescales longer than $B$, and declines for shorter
 timescales. Each power law corresponds to a separate class of physical phenomena with a different characteristic timescale. The fitting of
solar data \citep{2004A&A...414.1139A} uncovers three components. The first component corresponds to active regions 
($\tau \simeq$ 1.3 x 10$^5$ s), whose amplitude both increases and correlates with the \ion{Ca}{ii} K-line index 
\citep{1994SoPh..152..247A,2004A&A...414.1139A}. The second component is related to super- and meso-granulation
with typical timescales of hours, but no detailed models of these phenomena have been developed to date. The third component is the superposition of
variability on timescales of a few minutes, related to granulation and higher frequency effects, such as oscillations and photon noise.

This model can be applied to other stars \citep{2004A&A...414.1139A} to predict the expected ``stellar background''. According to
\citet{2004A&A...414.1139A}, the amplitude of the first power law A$_1$
is correlated with emission in the \ion{Ca}{ii} H \& K lines, $i$.$e$. $R'_{\rm HK}$, and can be written as follows
\begin{equation}
A_{\rm 1} = 2.20 \times 10^{\rm -5} + 3.04 R'_{\rm HK} + 1.90 (R'_{\rm HK})^2 \times 10^{\rm 5}.
\end{equation}
We refer the reader to \citet{2004A&A...414.1139A} for a detailed derivation of the aforementioned formula and the other two
parameters ($B_{\rm i}$ and $C_{\rm i}$). Chromospheric activity measurements, such as those presented in this work, can be therefore used 
as a {\itshape proxy} to infer the expected amplitude variation of active stars and thus to establish a lower limit to planet detection.

\section{Summary and conclusions}

We have used high resolution spectroscopic observations to measure the chromospheric activity and the projected rotational velocities for 
371 nearby cool stars. For the fraction presenting chromospheric activity (173 stars out of 371), we have analysed the 
relationship between pairs of chromospheric activity indicator lines, compiling empirical relations to be used when not all the chromospheric 
features are included in the spectral range. We have 
applied these relationships to obtain values of $\log R'_{\rm HK}$ when the \ion{Ca}{ii} H \& K spectral region was not available 
in the spectrum.

To test the applicability of the results to planet searches, we have calculated the $RV$ jitter one should expect for 
each of the active stars in the sample. As previously pointed out, those values must be applied carefully 
because magnetic activity is variable and a simple subtraction of the activity-related ``signal'' is not possible. 
They have to be used as an estimation of the activity-related noise one should expect for a star and thus used to 
set the minimum detectable mass for a planet orbiting the star or to determine the minimal amplitude variation that could 
indicate the existence of a planet. Our results represent an important resource in terms of target selection for exoplanet searches surveys.

\begin{acknowledgements}
R. Mart\'inez-Arn\'aiz acknowledges support from the Spanish Ministerio de Educaci\'on y Ciencia (currently the Ministerio de Ciencia e 
Innovaci\'on), under the grant FPI20061465-00592 (programa nacional Formaci\'on Personal Investigador) and projects AYA2008-00695 
(Programa Nacional de Astronom\'ia y Astrof\'isica), AYA2008-01727 (Programa Nacional de Astronom\'ia y Astrof\'isica), AstroMadrid
S2009/ESP-1496. This research has made use of the SIMBAD database and VizieR catalogue access 
tool, operated at CDS, Strasbourg, France. We also thank the anonymous referee for his/her valuable suggestions on how to improve 
the manuscript.
\end{acknowledgements}

\bibliographystyle{aa}
\bibliography{activity_rma_final}

\Online

\appendix 
\section{Tables of results}

The stellar and line parameters are published in electronic format only. available at CDS, Table \ref{parameters}, contains the 
Hipparcos number (Col. \#1), the spectrograph used to observe the star (Col. \#2), the modified julian date (MJD) of the observation 
(Col. \#3), the right ascension and declination (Col. \#4 and \#5), colour index ($B$--$V$) (Col. \#6), spectral type (Col. \#7), 
and projected rotational velocity, $v$~$\sin{i}$ (Col. \#8). Col. \#9 specifies whether the star may be classified as active or non active. 
We note that for some stars in Col. \#8, only upper limits are given. As mentioned in the text, for very slowly rotating stars, the value of 
$\sigma_{\rm 0}$ can be higher than that of $\sigma_{\rm obs}$. In those cases, we give the minimum value that could be measured with 
the same spectrograph and for a star of the same spectral type.\\

The chromospheric activity results are listed in two different tables. Table \ref{tab:activity_ew} contains the excess emission EW as 
measured in the subtracted spectrum, whereas Table \ref{tab:activity_flux} includes the excess fluxes derived in this work. In both tables,
Cols. \#1 , \#2, and \#3 contain the Hipparcos number of the star, the spectrograph used to observe it, and the modified julian date of 
the observation, respectively. In Cols. \#4, \#5, \#6, \#7, \#8, and \#9, excess emission (or fluxes) for  Ca \textsc{ii} K, Ca \textsc{ii} H, 
H$\alpha$, and Ca \textsc{ii} IRT $\lambda$4898\AA, Ca \textsc{ii} IRT $\lambda$8542\AA\, and Ca \textsc{ii} IRT $\lambda$8662\AA\, are given. 
Table \ref{tab:activity_flux} has an additional column containing $\log R'_{\rm HK}$.
 As mentioned in the text, for those stars with measured values of both Ca \textsc{ii} H and Ca \textsc{ii} K lines, $\log R'_{\rm HK}$ was derived 
directly as described in Sect. \ref{s_Rhk}. When it was not possble to measure one or both of the Ca \textsc{ii} lines, we used the empirical 
relationships between the total flux in Ca \textsc{ii} (H + K) and H$\alpha$ or Ca \textsc{ii} IRT obtained in the present work 
(see Fig. \ref{flux} and Table \ref{tab:flux_rel}).
Given that the relationship between Ca \textsc{ii} (H + K) and Ca \textsc{ii} IRT clearly exhibits lower dispersion, we used it 
when possible, $i.e$ when measuring the three lines 
in the Ca \textsc{ii} infrared triplet was possible. In the remaining cases, including all the stars observed with SARG 
(infrared orders are strongly affected by fringing), we used the relationship between Ca \textsc{ii} (H + K) and H$\alpha$.\\

Predicted radial velocity variations, $i$.$e.$, jitter \citep[based on][]{1998ApJ...498L.153S,2000A&A...361..265S}, are given in Table \ref{jitter}. 
It contains only those stars for which $\log R'_{\rm HK}$ could be derived. Columns \#1, \#2, and \#3  contain the same information as that of Tables 
\ref{tab:activity_ew} and \ref{tab:activity_flux}. The spectral type and $\log R'_{\rm HK}$ for each star are listed in Cols. \#4 and \#5. 
Columns \#6 and \#7 contain the $\sigma_{\rm rv}$ values (within 1$\sigma$) obtained using \citet{1998ApJ...498L.153S} and \citet{2000A&A...361..265S} 
relationships, respectively.\\

\begin{figure}
\centering
\includegraphics[width=8cm, keepaspectratio]{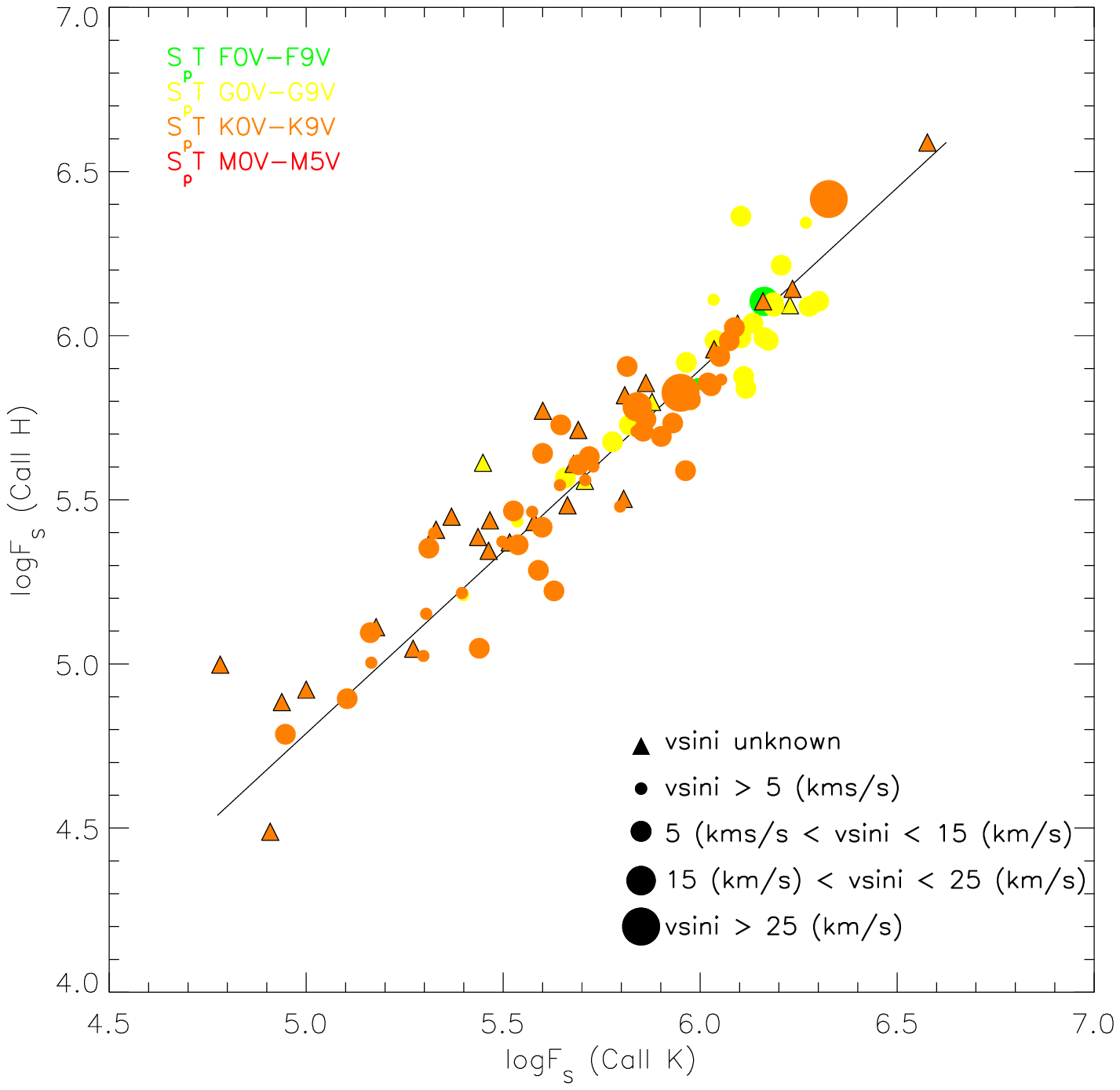}
\hfill
\includegraphics[width=8cm, keepaspectratio]{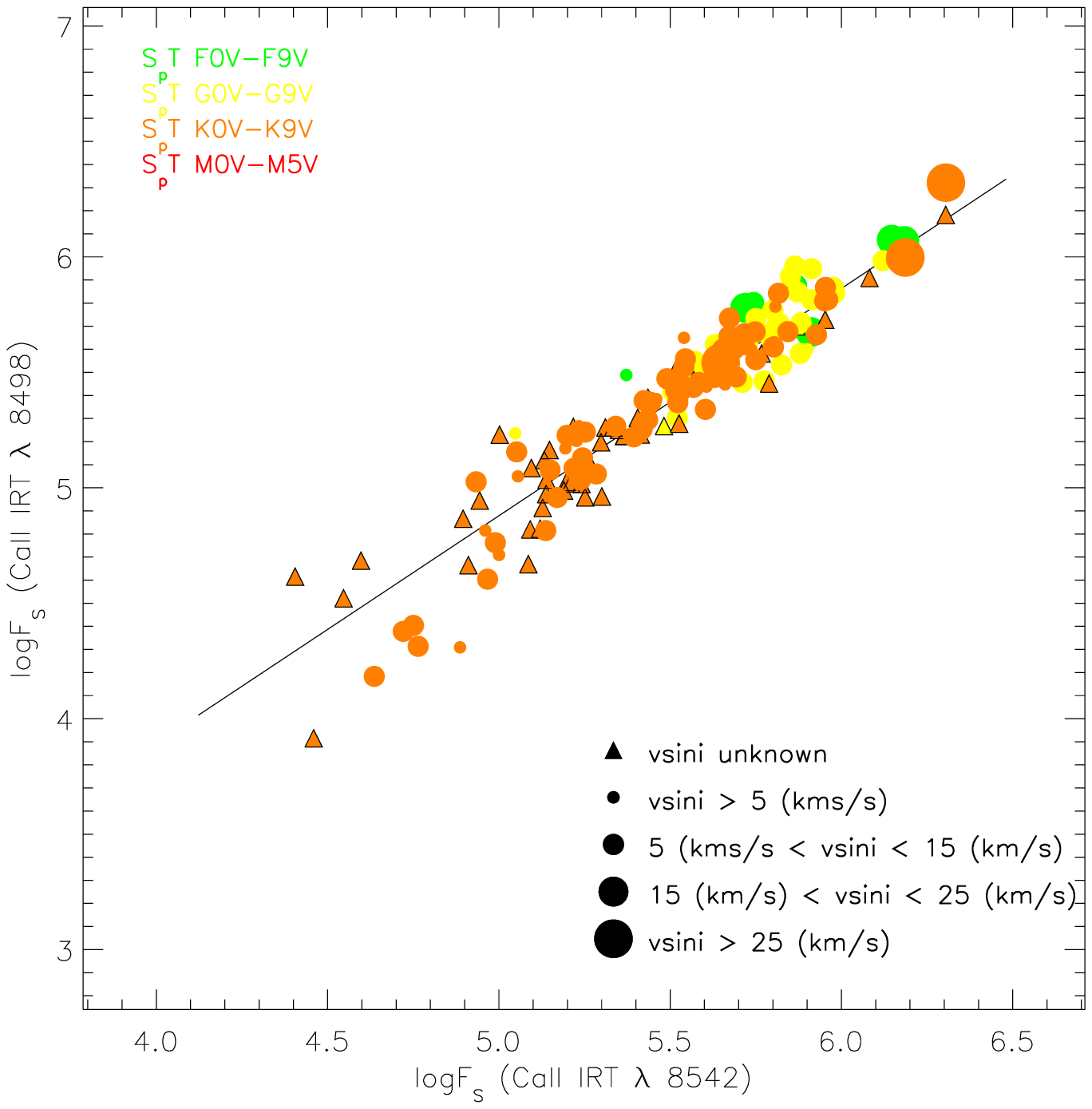}
\vfill
\includegraphics[width=8cm, keepaspectratio]{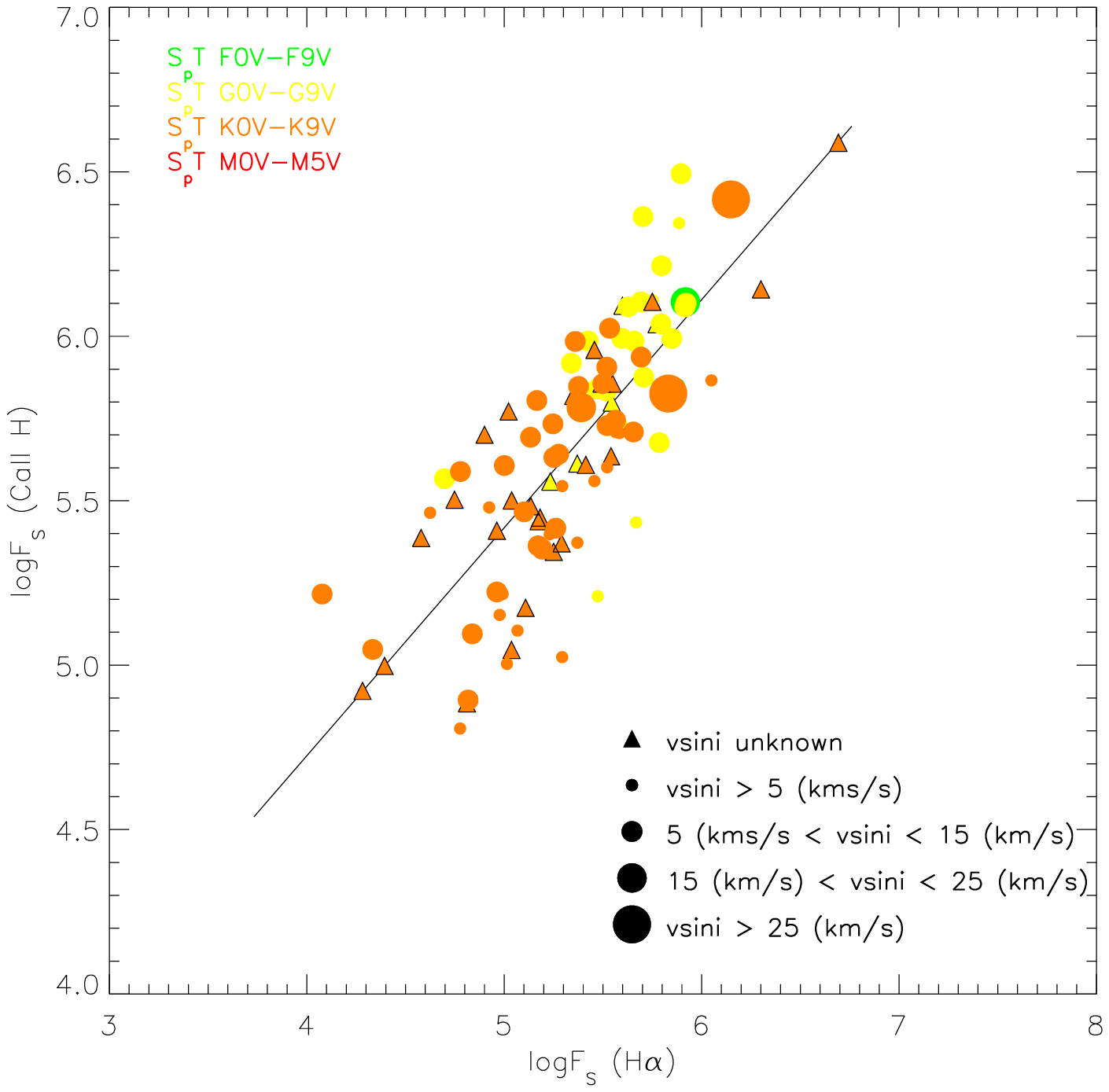}
\hfill
\includegraphics[width=8cm, keepaspectratio]{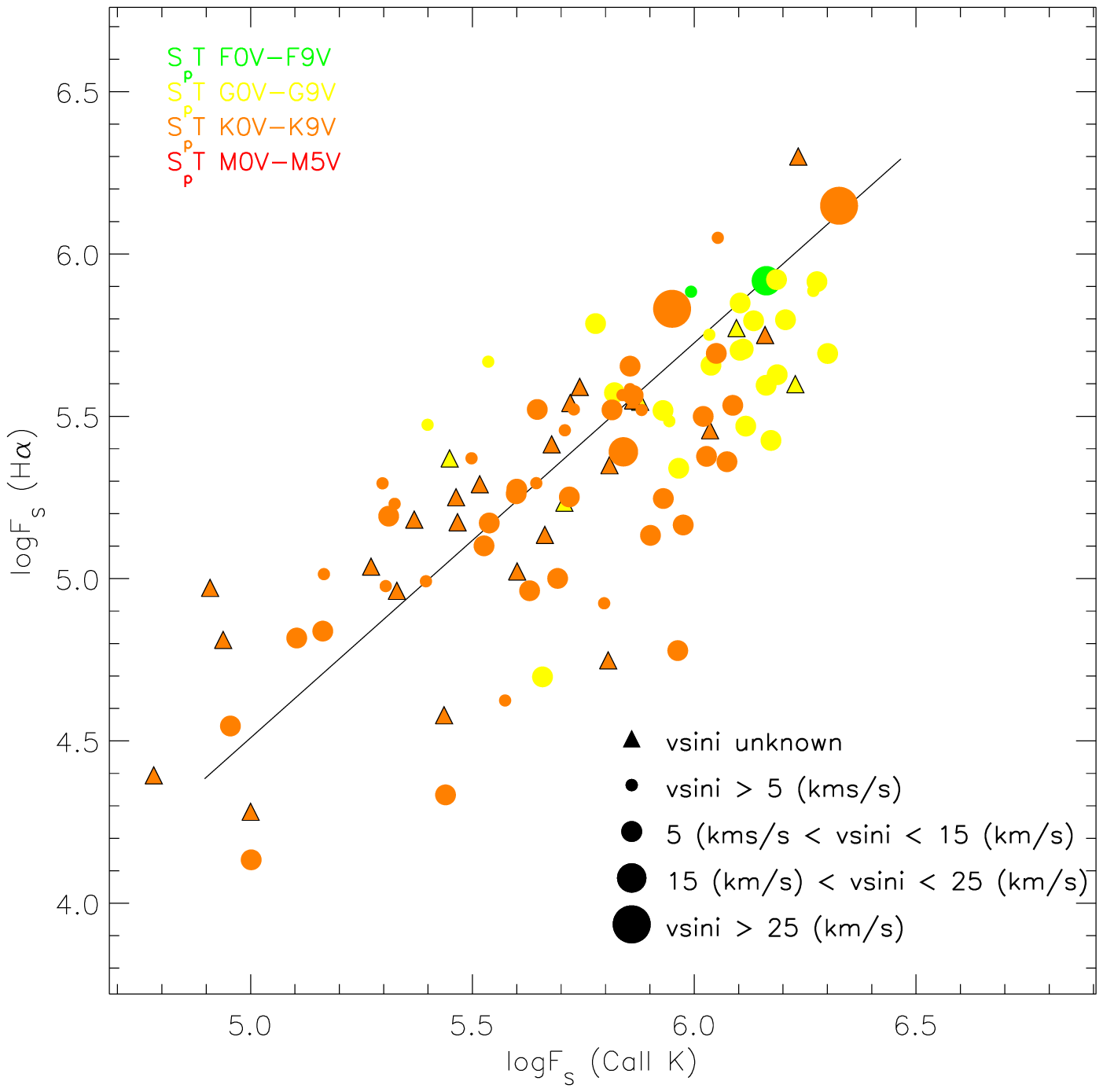}
\vfill
\includegraphics[width=8cm, keepaspectratio]{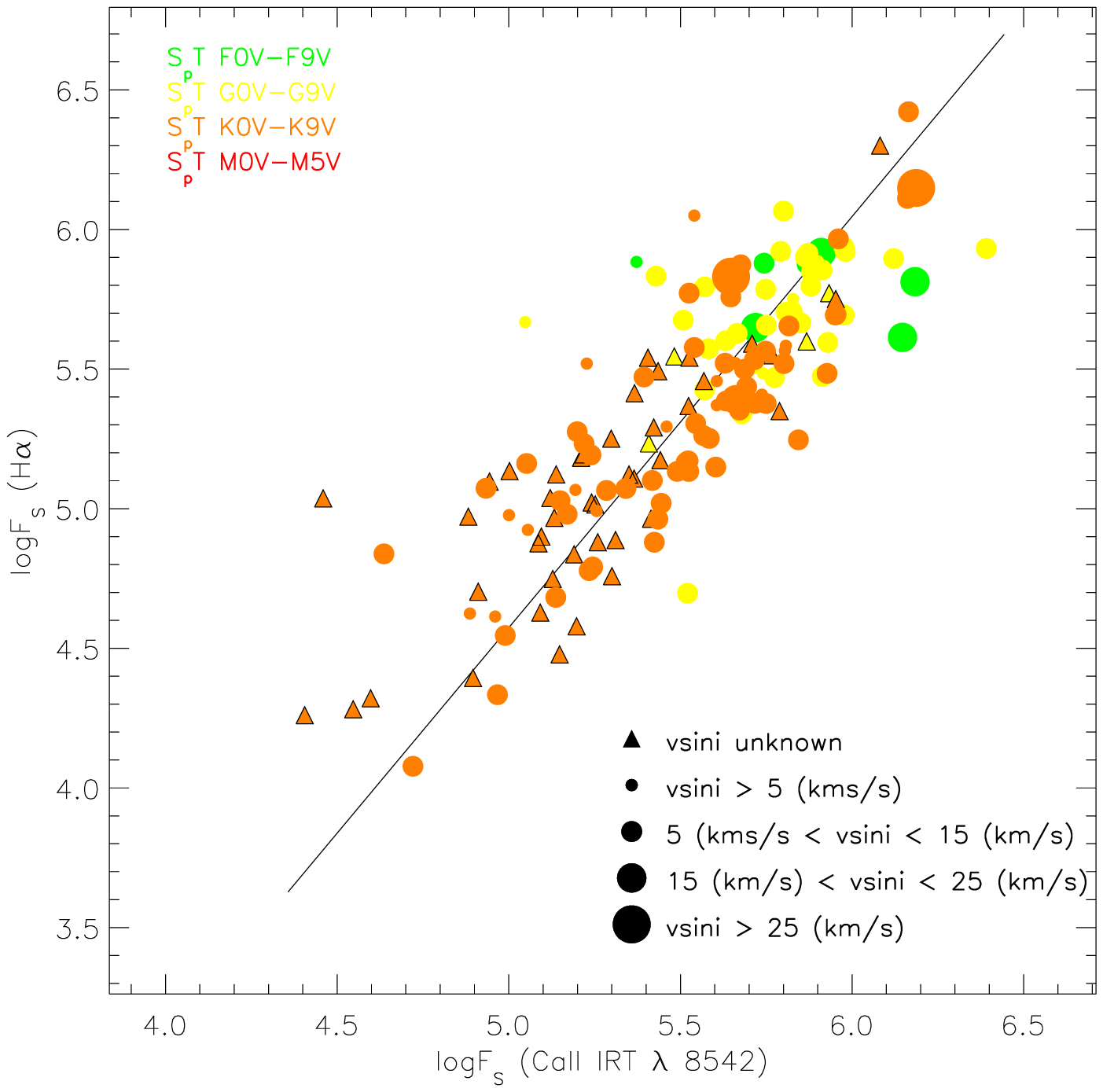}
\hfill
\includegraphics[width=8cm, keepaspectratio]{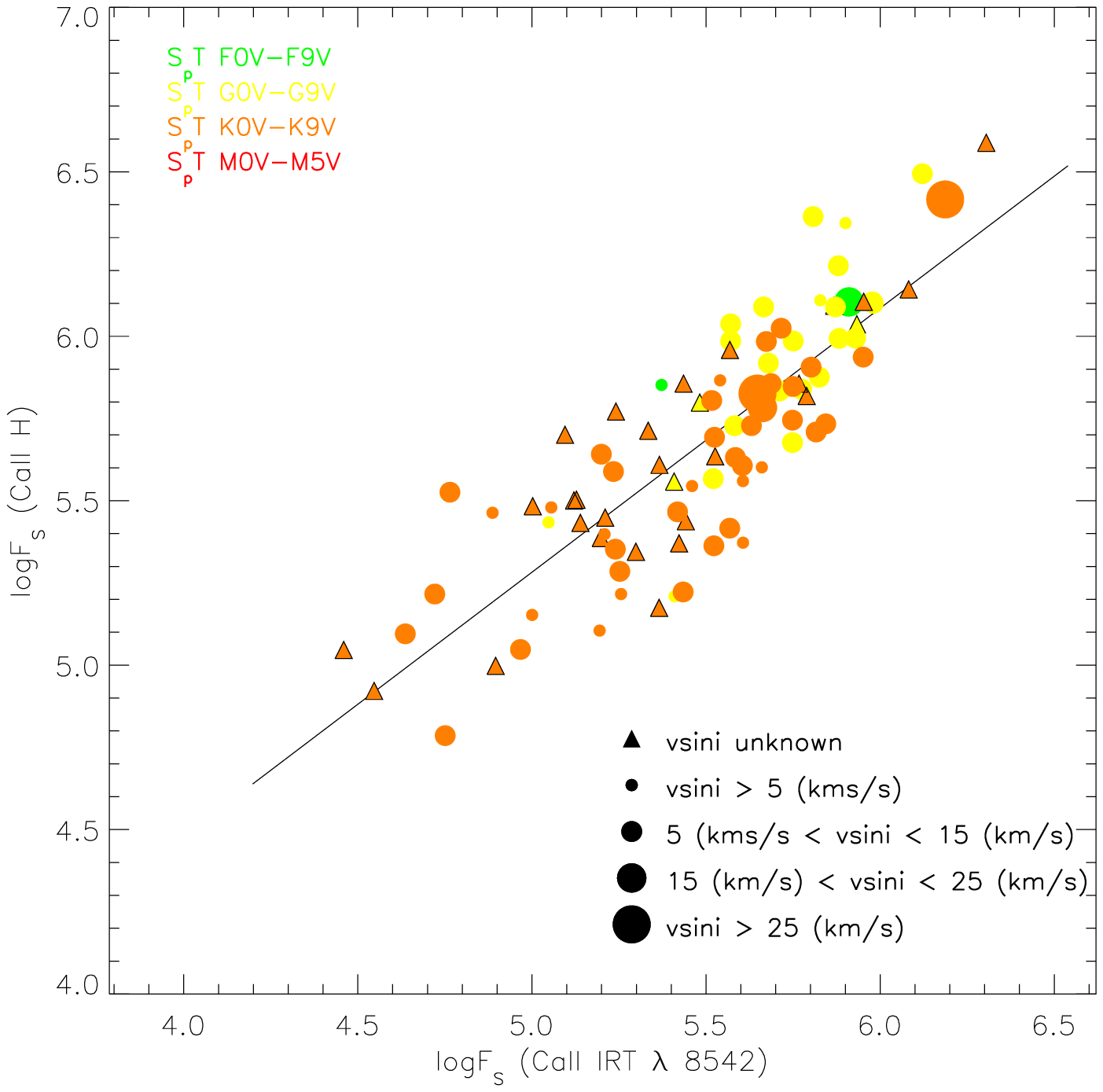}
\caption[Flux-flux relationships among different chromospheric activity indicators]
{Flux-flux relationships among different chromospheric activity indicators. Symbol thickness increases with increasing rotational velocity
(triangles are used when $v$~$\sin{i}$ could not be determined). Colors are used to discern different spectral types.}
\label{flux_all}
\end{figure}

As online material we have also included Fig. \ref{flux_all} with plots of the flux-flux relationships among different chromospheric activity indicators. 

\tiny
\begin{longtable}{l l c l r c l r l}
\caption[]{Stellar parameters of the stars.
\label{parameters}} \\
\hline
\hline
\noalign{\smallskip}
HIP  & Spectrograph & MJD & \multicolumn{1}{c}{RA} & \multicolumn{1}{c}{DEC} &
        \multicolumn{1}{c}{$B-V$} & \multicolumn{1}{l}{SpT} & \multicolumn{1}{c}{$v \sin i$} & ACTIVE \\
      \noalign{\smallskip}
    &  & \multicolumn{1}{c}{(days)} & \multicolumn{1}{c}{(J2000)} & \multicolumn{1}{c}{(J2000)} &
            \multicolumn{1}{c}{(mag)} &        &   (km\,s$^{\rm -1}$) & \\
\noalign{\smallskip}
\hline
\noalign{\smallskip}
\endfirsthead
\caption{continue} \\
\hline
\hline
\noalign{\smallskip}
HIP  & Spectrograph & MJD & \multicolumn{1}{c}{RA} & \multicolumn{1}{c}{DEC} &
        \multicolumn{1}{c}{$B-V$} & \multicolumn{1}{l}{SpT} & \multicolumn{1}{c}{$v \sin i$}  & ACTIVE \\
      \noalign{\smallskip}
    &  & \multicolumn{1}{c}{(days)} & \multicolumn{1}{c}{(J2000)} & \multicolumn{1}{c}{(J2000)} &
        \multicolumn{1}{c}{(mag)} &        &   (km\,s$^{\rm -1}$)  & \\
\noalign{\smallskip}
\hline
\noalign{\smallskip}
\endhead
\hline
\multicolumn{2}{l}{\dag\, stars with known exoplanets}\\
\multicolumn{6}{l}{$\star$ stars with chromospheric variations during our observations}\\
\multicolumn{9}{l}{* stars classified as active by visual inspection but for which chromospheric activity could not be measured}\\
%\multicolumn{8}{l}{  for which chromospheric activity could not be measured}
\endfoot
\hline
\multicolumn{2}{l}{\dag\, stars with known exoplanets}\\
\multicolumn{6}{l}{$\star$ stars with chromospheric variations during our observations}\\
\multicolumn{9}{l}{* stars classified as active by visual inspection but for which chromospheric activity could not be measured}\\
%\multicolumn{8}{l}{  for which chromospheric activity could not be measured}
\endlastfoot
171	&	McDonald	&	52164.3858	&	00 02 09.65	&	+27 05 04.2	&	0.690	&	G5Vb	&	 $\leq$ 4.07	&	Not Active	\\
544	&	McDonald	&	52031.3248	&	00 06 36.53	&	+29 01 19.0	&	0.749	&	G8V	&	5.72	&	Active	\\
544	&	SARG	&	54780.9561	&	00 06 36.53	&	+29 01 19.0	&	0.749	&	G8V	&	8.07	&	Active	\\
910	&	SARG	&	54777.8430	&	00 11 15.91	&	-15 28 02.4	&	0.489	&	F5V	&	4.88	&	Not Active	\\
1499\dag &	FOCES	&	53578.1286	&	00 18 41.62	&	-08 03 09.5	&	0.680	&	G1.5V	&	4.18	&	Not Active	\\
1532	&	FOCES	&	54089.7731	&	00 19 05.58	&	-09 57 50.8	&	1.305	&	K5V	&	 $\leq$ 0.53	&	Not Active	\\
1598	&	FOCES	&	53576.1488	&	00 20 00.51	&	+38 13 41.0	&	0.625	&	G0V	&	3.74	&	Not Active	\\
1599	&	FEROS	&	52030.4120	&	00 20 01.91	&	-64 52 39.4	&	0.576	&	F9V	&	 $\leq$ 3.23	&	Not Active	\\
\end{longtable}

\normalsize
%
%\newpage
%\section{Chromospheric activity}
%The chromospheric activity data are published in electronic format only. We have separated the data into two different tables. The former contains 
%the excess emission EW as measured in the subtrated spectrum, whereas in the latter we include the excess fluxes derived in this work. In both tables, 
%column \#1 and \#2 contain the Hipparcos name of the star and the spectrograph used to observe it, respectively. In columns \#3, \#4, \#5, \#6, \#7 
%and \#8, excess emission (or fluxes) for  Ca \textsc{ii} K & Ca \textsc{ii} H, H$\alpha$ and Ca \textsc{ii} IRT $\lambda$4898\AA, 
%Ca \textsc{ii} IRT $\lambda$8542\AA) and Ca \textsc{ii} IRT $\lambda$8662\AA. Table 2 presents an additional column containing $\log R'_{HK}$.
%
\tiny
\begin{longtable}{ l l c c c c c c c }
\caption{Excess emission in different chromospheric activity indicator lines for the active stars in the sample.
\label{tab:activity_ew}} \\
\hline
\hline
\noalign{\smallskip}
& & & \multicolumn{6}{c}{EW (\AA) in the subtrated spectrum}\\ \cline{4-9}
{HIP} & {Spectrograph} & {MJD} & \multicolumn{2}{c}{Ca \textsc {ii}}
 & & \multicolumn{3}{c}{Ca \textsc {ii} IRT}\\
\cline{4-5} \cline{7-9} & & (days)  & {K} & {H} & ${H\alpha}$ & ${\lambda8498}$ & ${\lambda8542}$ & ${\lambda8662}$\\
\noalign{\smallskip}
\hline
\noalign{\smallskip}
\endfirsthead
\caption{continue} \\
\hline
\hline
\noalign{\smallskip}
& & & \multicolumn{6}{c}{EW (\AA) in the subtrated spectrum}\\ \cline{4-9}
{HIP} & {Spectrograph} & {MJD} & \multicolumn{2}{c}{Ca \textsc {ii}}
 & & \multicolumn{3}{c}{Ca \textsc {ii} IRT}\\
\cline{4-5} \cline{7-9} & & (days) & {K} & {H} & ${H\alpha}$ & ${\lambda8498}$ & ${\lambda8542}$ & ${\lambda8662}$\\
\noalign{\smallskip}
\hline
\noalign{\smallskip}
\endhead
\hline
\endfoot
\hline
\endlastfoot
544 & SARG & 54780.9561 & ... & ... & 0.056 $\pm$ 0.016 & 0.239 $\pm$ 0.067 & 0.220 $\pm$ 0.105 & 0.228 $\pm$ 0.084 \\
544 & McDonald & 52031.3248 & 0.112 $\pm$ 0.008 & 0.089 $\pm$ 0.008 & 0.115 $\pm$ 0.036 & 0.133 $\pm$ 0.023 & 0.151 $\pm$ 0.020 & 0.164 $\pm$ 0.006 \\
1803 & SARG & 54777.9856 & ... & ... & 0.135 $\pm$ 0.013 & 0.177 $\pm$ 0.081 & 0.227 $\pm$ 0.066 & ... \\
3093 & FOCES & 53577.1854 & 0.062 $\pm$ 0.004 & ...  & ...  & ...  & ...  & ...  \\
3765 & FOCES & 53747.7686 & 0.035 $\pm$ 0.030 & 0.319 $\pm$ 0.019 & 0.004 $\pm$ 0.011 & ... & ... & ...\\
3765 & McDonald & 52164.3977 & 0.032 $\pm$ 0.006 & 0.023 $\pm$ 0.006 & 0.009 $\pm$ 0.025 & 0.019 $\pm$ 0.010 & 0.033 $\pm$ 0.003 & 0.023 $\pm$ 0.006 \\
4148 & FEROS & 51834.1339 & 0.124 $\pm$ 0.007 & 0.093 $\pm$ 0.009 & 0.066 $\pm$ 0.015 & 0.095 $\pm$ 0.013 & 0.140 $\pm$ 0.014 & 0.087 $\pm$ 0.006 \\
4845 & SARG & 54778.9878 & ... & ... & 0.073 $\pm$ 0.017 & 0.050 $\pm$ 0.020 & 0.073 $\pm$ 0.101 & 0.063 $\pm$ 0.087 \\
\end{longtable}

\normalsize
\tiny
\begin{longtable}{ l l c c c c c c c c }
\caption{Excess flux in different chromospheric activity indicator lines for the active stars in the sample.
\label{tab:activity_flux}} \\
\hline
\hline
\noalign{\smallskip}
& & & \multicolumn{6}{c}{log$F_{\rm S}$ (erg~cm$^{\rm -2}$~s$^{\rm -1}$) in the subtrated spectrum} & \\ \cline{4-9}
{HIP} & {Spectrograph} & {MJD} & \multicolumn{2}{c}{Ca \textsc {ii}}
 & & \multicolumn{3}{c}{Ca \textsc {ii} IRT} &  {log$R'_{\rm HK}$}\\
\cline{4-5} \cline{7-9} & & (days) & {K} & {H} & ${H\alpha}$ & ${\lambda8498}$ & ${\lambda8542}$ & ${\lambda8662}$ & \\
\noalign{\smallskip}
\hline
\noalign{\smallskip}
\endfirsthead
\caption{continue} \\
\hline
\hline
\noalign{\smallskip}
& & & \multicolumn{6}{c}{log$F_{\rm S}$ (erg~cm$^{\rm -2}$~s$^{\rm -1}$) in the subtrated spectrum} &\\ \cline{4-9}
{HIP} & {Spectrograph} &{MJD} &  \multicolumn{2}{c}{Ca \textsc {ii}}
 & & \multicolumn{3}{c}{Ca \textsc {ii} IRT} & {log$R'_{\rm HK}$}\\
\cline{4-5} \cline{7-9} & & (days) & {K} & {H} & ${H\alpha}$ & ${\lambda8498}$ & ${\lambda8542}$ & ${\lambda8662}$ & \\
\noalign{\smallskip}
\hline
\noalign{\smallskip}
\endhead
\hline
\endfoot
\hline
\endlastfoot
544 & SARG & 54780.9561 & ... & ... &  5.47 $\pm$  0.12 &  5.95 $\pm$  0.12 &  5.91 $\pm$  0.21 &  5.93 $\pm$  0.16 &   -4.61 \\
544 & McDonald & 52031.3248 &  5.78 $\pm$  0.03 &  5.68 $\pm$  0.04 &  5.79 $\pm$  0.14 &  5.69 $\pm$  0.08 &  5.75 $\pm$  0.06 &  5.79 $\pm$  0.02 &   -4.67 \\
1803 & SARG & 54777.9856 & ... & ... &  5.94 $\pm$  0.04 &  5.87 $\pm$  0.20 &  5.98 $\pm$  0.13 & ... &   -4.36 \\
3093 & FOCES & 53577.1854 &  5.32 $\pm$  0.03 & ... & ... & ... & ... & ... & ... \\
3765 & FOCES & 53747.7686 &  5.00 $\pm$  0.37 &  5.96 $\pm$  0.03 &  4.13 $\pm$  1.38 & ... & ... & ... &   -5.41 \\
3765 & McDonald & 52164.3977 &  4.95 $\pm$  0.08 &  4.81 $\pm$  0.12 &  4.55 $\pm$  1.18 &  4.76 $\pm$  0.22 &  4.99 $\pm$  0.04 &  4.83 $\pm$  0.12 &   -5.41 \\
4148 & FEROS & 51834.1339 &  5.50 $\pm$  0.03 &  5.37 $\pm$  0.04 &  5.37 $\pm$  0.10 &  5.44 $\pm$  0.06 &  5.61 $\pm$  0.04 &  5.40 $\pm$  0.03 &   -4.83 \\
4845 & SARG & 54778.9878 & ... & ... &  5.12 $\pm$  0.10 &  4.97 $\pm$  0.17 &  5.14 $\pm$  0.60 &  5.07 $\pm$  0.60 &   -4.54 \\
\end{longtable}

\normalsize
\tiny
\begin{longtable}{l l c l l l l}
\caption[]{Predicted radial velocity jitter.
\label{jitter}}
\renewcommand{\footnoterule}{} \\
\hline
\hline
\noalign{\smallskip}
{HIP} & {Spectrograph} & {MJD} & {SpT} & {log$R'_{\rm HK}$} &
{$\sigma_{\rm rv}$\footnotemark[1]} & 
{$\sigma_{\rm rv}$\footnotemark[2]}\\
& &  (days) & & & (m\,s$^{\rm -1}$) & (m\,s$^{\rm -1}$)\\
\noalign{\smallskip}
\hline
\noalign{\smallskip}
\endfirsthead
\caption{continue} \\
\hline
\hline
\noalign{\smallskip}
{HIP} & {Spectrograph} & {MJD} & {SpT} & {log$R'_{\rm HK}$} &
{$\sigma_{\rm rv}$\footnotemark[1]} & 
{$\sigma_{\rm rv}$\footnotemark[2]}\\
& &  (days) & & & (m\,s$^{\rm -1}$) & (m\,s$^{\rm -1}$)\\
\noalign{\smallskip}
\hline
\noalign{\smallskip}
\endhead
\hline
\multicolumn{2}{l}{\footnotemark[1] \citet{1998ApJ...498L.153S}} & \multicolumn{5}{l}{\ddag Stars with  $\log R'_{\rm HK} \le -4.9$}\\
\multicolumn{2}{l}{\footnotemark[2] \citet{2000A&A...361..265S}}
\endfoot
\hline
\multicolumn{2}{l}{\footnotemark[1] \citet{1998ApJ...498L.153S}} & \multicolumn{5}{l}{\ddag Stars with  $\log R'_{\rm HK} \le -4.9$}\\
\multicolumn{2}{l}{\footnotemark[2] \citet{2000A&A...361..265S}}
\endlastfoot
544 & SARG & 54780.9561 & G8V &   -4.61 &   5 -  18 &   8 -  19 \\
544 & McDonald & 52031.3248 & G8V &   -4.67 &   5 -  15 &   7 -  18 \\
1803 & SARG & 54777.9856 & G3V &   -4.36 &  11 -  34 &  11 -  26 \\
3765\ddag & FOCES & 53747.7686 & K2V &   -5.41 &   0 -   2 &   4 -  10 \\
3765\ddag & McDonald & 52164.3977 & K2V &   -5.41 &   0 -   2 &   4 -  10 \\
4148 & FEROS & 51834.1339 & K2V &   -4.83 &   3 -  10 &   5 -  12 \\
4845 & SARG & 54778.9878 & K7V &   -4.54 &   6 -  21 &   5 -  13 \\
5286 & FOCES & 53576.1601 & K3V &   -4.59 &   6 -  19 &   5 -  13 \\
\end{longtable}

\normalsize

\onecolumn
\twocolumn

%\include{}
%\label{tab:activity_ew}
%\include{tab:activity_flux}
%\label{tab:activity_flux}

\end{document}